\documentclass[prd,aps,superscriptaddress,11pt,nofootinbib]{revtex4-2}
\usepackage[utf8]{inputenc}
\usepackage{graphicx}
\usepackage{color}
\usepackage{babel}
\usepackage{array}
\usepackage{multirow}
\usepackage{amsmath}
\usepackage{amssymb}
\usepackage{graphicx}
\usepackage[unicode=true,pdfusetitle,
 bookmarks=true,bookmarksnumbered=false,bookmarksopen=false,
 breaklinks=false,pdfborder={0 0 0},pdfborderstyle={},backref=false,colorlinks=true]
 {hyperref}
\hypersetup{
 linkcolor=blue, citecolor=cyan}

\makeatletter

\providecommand{\tabularnewline}{\\}

\usepackage{amsthm}

\usepackage{xcolor}

\allowdisplaybreaks[2]

\begin{document}
%\title{Attractors in the relativistic spin hydrodynamics with Gubser flow}
\title{Late-time attractors in relativistic spin hydrodynamics in Gubser flow}

\author{Gen-Hui Li}
\email{genhuili@mail.ustc.edu.cn}
\affiliation{Department of Modern Physics and Anhui Center for Fundamental Sciences in Theoretical Physics, University of Science and Technology of China, Anhui 230026, China}

\author{Xiang Ren}
\email{renxiang0510@mail.ustc.edu.cn}
\affiliation{Department of Modern Physics and Anhui Center for Fundamental Sciences in Theoretical Physics, University of Science and Technology of China, Anhui 230026, China}

\author{Dong-Lin Wang}
\email{donglinwang@mail.ustc.edu.cn}
\affiliation{Department of Modern Physics and Anhui Center for Fundamental Sciences in Theoretical Physics, University of Science and Technology of China, Anhui 230026, China}
\affiliation{Department of Physics, The University of Tokyo, 
  7-3-1 Hongo, Bunkyo-ku, Tokyo 113-0033, Japan}

\author{Shi Pu}
\email{shipu@ustc.edu.cn}
\affiliation{Department of Modern Physics and Anhui Center for Fundamental Sciences in Theoretical Physics, University of Science and Technology of China, Anhui 230026, China}
\affiliation{Southern Center for Nuclear-Science Theory (SCNT), Institute of Modern Physics, Chinese Academy of Sciences, Huizhou 516000, Guangdong Province, China}

\begin{abstract}
We investigate the late-time asymptotic solutions and attractor structure of the spin density in minimal causal spin hydrodynamics in Gubser flow. After deriving the differential equation governing the spin density, we obtain its late-time asymptotic solutions and identify both attractors and repellers in the corresponding numerical solutions. We then map these solutions back to flat Minkowski space and find parameter regions where the spin density exhibits a power-law decay. We further show that, when the characteristic length scale of the system is much larger than the proper time, several components of the spin density can decay as slowly as conventional thermodynamic variables in relativistic hydrodynamics. In this regime, the spin density behaves as a hydrodynamic mode governed by the late-time scaling laws of the flow.
\end{abstract}

\maketitle

\section{Introduction}

%(1) Spin polarization in HIC

%(2) Puzzles: local polarization, shear induced polarization
%\citep{Liu:2021uhn,Fu:2021pok,Becattini:2021suc,Becattini:2021iol,Yi:2021ryh,Yi:2021unq,Florkowski:2021xvy,Alzhrani:2022dpi,Wu:2022mkr,Yi:2023tgg,Wu:2023tku} as well as collisional \cite{Sheng:2021kfc,Wang:2021qnt,Weickgenannt:2021cuo,Fang:2022ttm}

%(3) QKT \cite{Copinger:2025ovz}

%(4) Spin hydrodynamics

%(5) Problem: spin decays much faster than other quantities

%(6) Review of our previous work in Bjorken

%(7) Motivation of this work

%(8) Structure of this paper

In noncentral relativistic heavy-ion collisions, the colliding nuclei carry a large initial orbital angular momentum. Part of this angular momentum can be transferred to the spin degrees of freedom of produced hadrons, leading to spin polarization of hyperons and spin alignment of vector mesons through spin--orbit coupling \cite{Liang:2004ph,Liang:2004xn}. The RHIC-STAR Collaboration has measured the global polarization, i.e., the polarization along the direction of the initial orbital angular momentum, as a function of collision energy, for $\Lambda$ and $\overline{\Lambda}$, and other hyperons \cite{STAR:2017ckg,STAR:2018gyt,STAR:2020xbm,STAR:2021beb,STAR:2023nvo}. At high and intermediate collision energies, the observed global polarization is primarily driven by thermal vorticity and can be described by a variety of phenomenological models \cite{Karpenko:2016jyx,Li:2017slc,Sun:2017xhx,Wei:2018zfb,Vitiuk:2019rfv,Fu:2020oxj,Lei:2021mvp,Ryu:2021lnx}. The global polarization in low-energy collisions \cite{STAR:2021beb,STAR:2023nvo,HADES:2022enx}, in particular near and below the production threshold of the $\Lambda$ hyperon, has also been investigated \cite{Guo:2021udq, Deng:2020ygd, Deng:2021miw, Vitiuk:2019rfv,Sun:2025oib,Liu:2025kpp,Zheng:2025ngn,Xu:2026hxz}, but a definitive understanding has not yet been reached.

On the other hand, the local polarization of hyperons, namely the polarization along the beam direction and the out-of-plane direction as a function of azimuthal angle or transverse momentum, has not been successfully described for a long time by the same models that reproduce the global polarization. To address this discrepancy, a hydrodynamic effect referred to as shear-induced polarization \cite{Liu:2021uhn,Fu:2021pok,Becattini:2021suc,Becattini:2021iol,Hidaka:2017auj, Yi:2021ryh,Yi:2021unq,Florkowski:2021xvy,Alzhrani:2022dpi,Palermo:2022lvh,Buzzegoli:2022fxu,Wu:2022mkr,Yi:2023tgg,Wu:2023tku,Palermo:2024tza} has been identified as playing an important role in the local polarization. In addition, polarization induced by gradients of the baryon chemical potential (often discussed in terms of the spin Hall effect \cite{Fu:2022myl,Ivanov:2022geb, Hidaka:2017auj, Wu:2023tku} and anomalous spin Hall effect \cite{Fang:2024vds,Wang:2025mfz}), may become relevant at low collision energies. Partonic-interaction corrections have also been investigated systematically in recent studies, e.g. see Refs.~\cite{Fang:2022ttm, Fang:2024vds, Fang:2025pzy} and the references therein.
Nevertheless, achieving a quantitative description of local polarization at low energies remains challenging \cite{Yi:2021ryh,Yi:2023tgg,Wu:2023tku,Yi:2024kwu}. Recently, the CMS measurement of local polarization in p+Pb collisions \cite{CMS:2025nqr} has also been reported to be incompatible with standard modeling assumptions \cite{Yi:2024kwu, Yi:2025anh, Yi:2025sgk}. For further details, we refer to recent reviews \cite{Gao:2020vbh,Becattini:2022zvf,Becattini:2024uha,Niida:2024ntm} and the references therein.

To get a better understanding of spin polarization in relativistic heavy-ion collisions, it is essential to investigate the dynamical evolution of spin degrees of freedom in the medium produced in the collisions. One approach is quantum kinetic theory \cite{Gao:2012ix,Chen:2012ca,Hidaka:2016yjf,Hidaka:2017auj,Gao:2019znl,Weickgenannt:2019dks,Liu:2020flb,Weickgenannt:2020aaf,Weickgenannt:2021cuo,Sheng:2021kfc,Hidaka:2022dmn,Fang:2022ttm,Dong:2022yzt,Dong:2023cng,Fang:2023bbw,Yin:2024dnu,Fang:2024vds,Fang:2025pzy,Bhadury:2025boe}, which provides a microscopic effective description of the evolution of quasi-particles with spin. Another approach is to formulate macroscopic evolution equations for spin, namely relativistic spin hydrodynamics \cite{Montenegro:2017rbu,Florkowski:2017ruc,Florkowski:2018fap,Hattori:2019lfp,Fukushima:2020ucl}.
This motivates extending conventional relativistic hydrodynamics by incorporating spin as an additional hydrodynamic degree of freedom. Such formulations have been developed using quantum statistical field theory \cite{Becattini:2018duy,Hu:2022azy,Dey:2024cwo,Tiwari:2024trl,She:2024rnx}, quantum kinetic theory \cite{Florkowski:2018myy,Bhadury:2020puc,Shi:2020htn,Speranza:2020ilk,Bhadury:2020cop,Singh:2020rht,Peng:2021ago,Weickgenannt:2022zxs,Weickgenannt:2022jes,Weickgenannt:2022qvh,Bhadury:2022ulr,Wagner:2023cct,Florkowski:2024bfw,Wagner:2024fhf,Wagner:2024fry,Daher:2025pfq}, analyses based on the entropy principle \cite{Hattori:2019lfp, Fukushima:2020ucl, Gallegos:2020otk,Li:2020eon,Gallegos:2021bzp,She:2021lhe,Hongo:2021ona,Biswas:2023qsw,Fang:2025aig, Zhang:2026zee}, and the effective field-theory approach \cite{Montenegro:2017rbu,Montenegro:2017lvf}. Very recently, using spin hydrodynamics, an extended Bargmann-Michel-Telegdi equation that incorporates rotation and dissipative effects has been derived \cite{Fang:2025aig}. For further discussion of spin hydrodynamics, we refer to recent reviews \cite{Florkowski:2018fap,Shi:2023sxh,Huang:2024ffg} and the references therein.

Although the basic framework of relativistic spin hydrodynamics has been established, several open issues remain. One issue concerns causality and stability. It has been found that first-order spin hydrodynamics in a gradient expansion, formulated in the canonical form, is generically acausal and unstable \cite{Xie:2023gbo, Sarwar:2022yzs,Daher:2022wzf,Daher:2024bah}. Moreover, even after a minimal extension to second order, spin hydrodynamics can still exhibit unstable modes in a linear-mode analysis, even when the asymptotic causality conditions are satisfied \cite{Xie:2023gbo} \footnote{This behavior is markedly different from conventional relativistic hydrodynamics without spin, see, e.g., the discussion in Refs.~\cite{Denicol:2008ha,Pu:2009fj,Kovtun:2019hdm}. Also see Refs.~\cite{Heller:2022ejw,Gavassino:2023myj,Gavassino:2023mad,Wang:2023csj,Hoult:2023clg} for recent discussions of improved causality and stability criteria in linear-mode analyses for relativistic many-body theories.}. To address this issue, Ref.~\cite{Ren:2024pur} applied the thermodynamic stability method \cite{Gavassino:2021kjm} and derived causality and stability conditions. 
A second issue concerns the choice of pseudo-gauge. In addition to the canonical form \cite{Hattori:2019lfp,Hongo:2021ona}, the Belinfante form \cite{Fukushima:2020ucl} and other pseudo-gauge choices \cite{Florkowski:2018fap,Speranza:2020ilk,Weickgenannt:2022zxs,Dey:2023hft} can also be used in constructing spin hydrodynamics. Although there has been recent progress in this direction \cite{Li:2020eon,Buzzegoli:2021wlg,Buzzegoli:2024mra,Becattini:2025twu}, the problem remains unsettled and should be addressed systematically in future work. A third issue is that several recent studies \cite{Becattini:2023ouz,Florkowski:2024bfw,Becattini:2025oyi} have reported that the thermodynamic relations commonly employed in spin hydrodynamics at finite spin density and spin chemical potential differ from those obtained in microscopic approaches. By contrast, other works have found consistency between spin-hydrodynamic thermodynamics and microscopic results by carefully accounting for rotational effects \cite{Ambrus:2025dca} or by introducing an appropriate pseudo-gauge transformation \cite{Armas:2026bmw}.

In addition, there is another long-standing issue in spin hydrodynamics, especially in the canonical formulation. In canonical spin hydrodynamics, the spin density is not a conserved quantity \cite{Hattori:2019lfp}, in contrast to the conserved charge (number) density. It is commonly expected that the spin density relaxes more rapidly than other macroscopic hydrodynamic fields, which would make its effects difficult to observe on the freeze-out hypersurface. This expectation is supported, for example, by analytic solutions of canonical spin hydrodynamics in Bjorken flow \cite{Wang:2021ngp} and Gubser flow \cite{Wang:2021wqq}.
Very recently, a work by some of us \cite{Wang:2024afv} has shown the existence of late-proper-time attractors \cite{Heller:2015dha,Romatschke:2017vte,Blaizot:2017ucy,Denicol:2017lxn,Spalinski:2017mel,Strickland:2018ayk,Blaizot:2019scw,Brewer:2019oha,Almaalol:2020rnu,Blaizot:2020gql,Heller:2020anv,Heller:2020uuy,Heller:2021oxl,Blaizot:2021cdv,Chen:2022ryi} in minimal causal relativistic spin hydrodynamics within Bjorken flow. In particular, nontrivial attractor solutions were found, accompanied by a parametrically slow decay of the spin density. This suggests that the spin density can persist to late proper times and may still influence observables on the freeze-out hypersurface.
On the other hand, Bjorken flow assumes boost invariance and neglects the transverse expansion of the system. A more realistic framework with radial expansion is therefore desirable. In this respect, Gubser flow \cite{Gubser:2010ze,Gubser:2010ui}, which incorporates transverse expansion while preserving boost invariance along the longitudinal direction, provides a natural extension.

This motivates the present work. Here we investigate the possible existence of attractor behavior for the spin density in canonical spin hydrodynamics under Gubser flow. We analyze the late-time asymptotic solutions for the spin density and identify the corresponding attractor and repeller branches for different parameter choices. Finally, we determine the conditions under which the spin density exhibits a power-law behavior at late times.

This paper is organized as follows. In Sec.~\ref{sec:spin_hydro}, we briefly review the minimal causal spin hydrodynamics in Gubser flow and derive the differential equation governing the spin density. In Sec.~\ref{sec:Asymptotic_solutions}, we derive the asymptotic late-time solutions for the spin density for different parameter choices. In Sec.~\ref{sec:attractor}, we present numerical results for the attractor and repeller branches. In Sec.~\ref{sec:S_late_time}, we discuss the late-time power-law decay of the spin density. Finally, we summarize in Sec.~\ref{sec:summary}.

We adopt the metric $g_{\mu\nu} = \mathrm{diag}\{+, -, -, -\}$ in a flat Minkowski spacetime with coordinates $(t,x,y,z)$ and the corresponding projector is defined as $\Delta_{\mu\nu}  =  g_{\mu\nu}-u_{\mu}u_{\nu}$ with $u^\mu$ being the fluid velocity. 
For an arbitrary tensor $A^{\mu\nu}$, we introduce the following convention for simplicity  $A^{(\mu\nu)} = \frac{1}{2}(A^{\mu\nu}+A^{\nu\mu})$, $A^{[\mu\nu]}  =  \frac{1}{2}(A^{\mu\nu}-A^{\nu\mu})$, and 
$A^{<\mu\nu>}  =  \frac{1}{2}(\Delta^{\mu\alpha}\Delta^{\nu\beta}+\Delta^{\nu\alpha}\Delta^{\mu\beta})A_{\alpha\beta}-\frac{1}{3}\Delta^{\mu\nu}(A^{\rho\sigma}\Delta_{\rho\sigma})$.
Throughout this work, we use standard notation for macroscopic quantities defined in flat Minkowski spacetime $\mathbb{R}^{3,1}$. For quantities defined in the $dS_3\times\mathbb{R}$ spacetime after the Weyl transformation, we add a hat. For example, $u^\mu$ and $\hat{u}^\mu$ denote the fluid velocity in $\mathbb{R}^{3,1}$ and in $dS_3\times\mathbb{R}$, respectively.

\section{Relativistic spin hydrodynamics with Gubser expansion}
\label{sec:spin_hydro}

We first briefly review the main equations of relativistic spin hydrodynamics in Sec.~\ref{subsec:spinhdyro}. We then review the standard Gubser-flow setup in Sec.~\ref{subsec:GubserFlow} and revisit spin hydrodynamics in $dS_3\times \mathbb{R}$ in Sec.~\ref{subsec:spin_dS3}.

\subsection{Brief review on relativistic spin hydrodynamics in canonical form} 
\label{subsec:spinhdyro}

We first review relativistic spin hydrodynamics in relativistic heavy-ion collisions. The main conservation laws are those of the energy-momentum tensor $\Theta^{\mu\nu}$ and the total angular momentum tensor $J^{\lambda\mu\nu}$,
\begin{eqnarray}
\nabla_{\mu}\Theta^{\mu\nu} & = & 0,\nonumber \\
\nabla_{\lambda}J^{\lambda\mu\nu} & = & 0.\label{eq:conservation_eq}
\end{eqnarray}
For simplicity, throughout this paper, we neglect the conservation equations for the conserved currents.
In this work, we follow \cite{Hattori:2019lfp, Fukushima:2020ucl} and adopt the relativistic spin hydrodynamics in canonical pesudo-gauge, in which the constitutive relations for $\Theta^{\mu\nu}$ and $J^{\lambda\mu\nu}$ take the form
\begin{eqnarray}
\Theta^{\mu\nu} & = & eu^{\mu}u^{\nu}-(p+\Pi)\Delta^{\mu\nu}+2h^{(\mu}u^{\nu)}+\pi^{\mu\nu}+2q^{[\mu}u^{\nu]}+\phi^{\mu\nu},\nonumber \\
J^{\lambda\mu\nu} & = & x^{\mu}\Theta^{\lambda\nu}-x^{\nu}\Theta^{\lambda\mu}+\Sigma^{\lambda\mu\nu},\label{eq:expression_J}
\end{eqnarray}
where $e$ and $p$ are the energy density and thermodynamic pressure, respectively, $u^{\mu}$ is the fluid four-velocity, and $\Sigma^{\lambda\mu\nu}$ is the rank-3 spin current tensor. Here, $\Pi$, $\pi^{\mu\nu}$, and $h^{\mu}$ denote the bulk viscous pressure, shear viscous tensor, and heat flux, respectively. For simplicity, we set the $h^\mu$, $\Pi$ and $\pi^{\mu\nu}$ to vanish in this work. 
The vector $q^\mu$ and the antisymmetric tensor $\phi^{\mu\nu}=-\phi^{\nu\mu}$ encode spin-related contributions. These quantities can be extracted from $\Theta^{\mu\nu}$ as
\begin{eqnarray}
h^{\mu}=u_{\nu}\Theta^{(\mu\nu)}, & \;& \pi^{\mu\nu}=\Theta^{<\mu\nu>},\nonumber \\
q^{\mu}=u_{\nu}\Theta^{[\mu\nu]}, & \;&
\phi^{\mu\nu}=\Delta^{\mu\alpha}\Delta^{\nu\beta}\Theta_{[\alpha\beta]}.
\end{eqnarray}
Substituting Eq.~\eqref{eq:expression_J} into Eq.~\eqref{eq:conservation_eq} yields
\begin{eqnarray}
\nabla_{\lambda}\Sigma^{\lambda\mu\nu} & = & -2\Theta^{[\mu\nu]}=-4q^{[\mu}u^{\nu]}-2\phi^{\mu\nu}.\label{eq:spin_tensor}
\end{eqnarray}
This evolution equation shows that the anti-symmetric part of the
energy-momentum tensor determines how spin evolves. We further decompose the spin current tensor $\Sigma^{\lambda\mu\nu}$ in analogy with the particle current,
\begin{eqnarray}
\Sigma^{\lambda\mu\nu} & = & u^{\lambda}S^{\mu\nu}+\Sigma_{(1)}^{\lambda\mu\nu}.
\end{eqnarray}
Here we have introduced the spin density tensor $S^{\mu\nu}=-S^{\nu\mu}$ and the residual contribution $\Sigma_{(1)}^{\lambda\mu\nu}$, which is orthogonal to $u^\mu$. Note that, in this work, $\Sigma^{\lambda\mu\nu}$ carries six independent degrees of freedom. One may alternatively consider a totally antisymmetric rank-3 spin tensor; see, e.g., \cite{Hongo:2021ona,Fang:2025aig}.
In analogy with the charge chemical potential, which is conjugate to the particle density, one can introduce a spin chemical potential $\omega_{\mu\nu}=-\omega_{\nu\mu}$ conjugate to the spin density $S^{\mu\nu}$. The thermodynamic relations of spin hydrodynamics then read \cite{Hattori:2019lfp, Fukushima:2020ucl},
\begin{eqnarray}
e+p & = & Ts+\omega_{\mu\nu}S^{\mu\nu},\nonumber \\
de & = & Tds+\omega_{\mu\nu}dS^{\mu\nu},\nonumber \\
dp & = & sdT+S^{\mu\nu}d\omega_{\mu\nu}. \label{eq:thermo_rel}
\end{eqnarray}
where $T$ and $s$ denote the temperature and entropy density, respectively. Recently, some works \cite{Becattini:2023ouz,Florkowski:2024bfw,Becattini:2025oyi} have reported that the thermodynamic relations derived from microscopic theory do not coincide with Eqs.~\eqref{eq:thermo_rel}. On the other hand, other studies \cite{Ambrus:2025dca,Armas:2026bmw} have found agreement between kinetic-theory results and Eqs.~\eqref{eq:thermo_rel}. The appropriate form of the thermodynamic relations in spin hydrodynamics therefore remains an open question. For simplicity, we adopt the conventional thermodynamic relations in this work.

Next, we introduce the power counting used in this work. We consider a gradient expansion in spin hydrodynamics and follow Ref.~\cite{Fukushima:2020ucl} by taking $\omega^{\mu\nu}\sim \mathcal{O}(\partial)$ and $S^{\mu\nu}\sim \mathcal{O}(1)$. It was shown in Refs.~\cite{Sarwar:2022yzs,Daher:2022wzf} that first-order spin hydrodynamics generally violates causality and stability. We therefore employ the \emph{minimal} causal theory proposed in Ref.~\cite{Xie:2023gbo}.
Here, \emph{minimal} means that we retain only the essential terms in the second-order gradient expansion required to ensure causality.
In this setup, the constitutive equations for $q^\mu$ and $\phi^{\mu\nu}$ read,
\begin{eqnarray}
\tau_{q}\Delta^{\mu\nu}u^{\alpha}\nabla_{\alpha}q_{\nu}+q^{\mu} & = & \lambda(T^{-1}\Delta^{\mu\alpha}\nabla_{\alpha}T+u^{\alpha}\nabla_{\alpha}u^{\mu}-4\omega^{\mu\nu}u_{\nu}),\nonumber \\
\tau_{\phi}\Delta^{\mu\alpha}\Delta^{\nu\beta}u^{\rho}\nabla_{\rho}\phi_{\alpha\beta}+\phi^{\mu\nu} & = & 2\gamma_{s}\Delta^{\mu\alpha}\Delta^{\nu\beta}(\nabla_{[\alpha}u_{\beta]}+2\omega_{\alpha\beta}),\label{eq:q_phi}
\end{eqnarray}
where $\lambda$ and $\gamma_{s}$ are two positive-definite transport coefficients, and $\tau_{q},\tau_{\phi}>0$ are the relaxation times for $q^\mu$ and $\phi^{\mu\nu}$, respectively.
When these transport coefficients satisfy the relations given in Ref.~\cite{Ren:2024pur}, the theory is causal and linearly stable.
For second-order spin hydrodynamics with additional dissipative terms, we refer to Refs.~\cite{Biswas:2023qsw,Daher:2024bah}.
Later, we will apply a Weyl transformation to Eq.~\eqref{eq:q_phi} and show that the constitutive equation for $\phi^{\mu\nu}$ breaks conformal symmetry. To restore conformal covariance, we introduce an additional term and write
\begin{eqnarray} \tau_{\phi}\Delta_{\alpha}^{\mu}\Delta_{\beta}^{\nu}u^{\rho}\nabla_{\rho}\phi^{\alpha\beta}+\lambda_\theta \phi^{\mu\nu}\nabla_{\alpha}u^{\alpha}+\phi^{\mu\nu}=2\gamma_{s}\Delta^{\mu\alpha}\Delta^{\nu\beta}(\nabla_{[\alpha}u_{\beta]}+2\omega_{\alpha\beta}),
\label{eq:con_phi}
\end{eqnarray}
where $\lambda_\theta$ is a transport coefficient that controls the strength of the coupling between $\phi^{\mu\nu}$ and the expansion rate $\nabla\cdot u$.
Such second-order coupling terms are generally allowed.

\subsection{Coordinates for Gubser flow}\label{subsec:GubserFlow}

In this section, we briefly review the conventional Gubser-flow setup introduced in Ref.~\cite{Gubser:2010ze,Gubser:2010ui}.
The Gubser flow is constructed to be invariant under the symmetry group $SO(3)\times SO(1,1)\times Z_{2}$.
Here, the $SO(1,1)$ factor corresponds to boost invariance, while the $Z_{2}$ factor corresponds to reflection symmetry across the collision plane.
The $SO(3)$ factor, in turn, encodes a specific combination of transverse translations and rotations in the transverse plane.
Because the fluid four-velocity is tightly constrained by these symmetries, it is nontrivial to construct a velocity profile that realizes them directly in Minkowski spacetime $\mathbb{R}^{3,1}$.
Exploiting conformal invariance, the dynamics are invariant under a Weyl rescaling of the metric, $g_{\mu\nu}\rightarrow\Lambda^{-2}g_{\mu\nu}$, where $\Lambda$ may vary across spacetime.
This allows one to realize the Gubser symmetry on a conformally related manifold and then map the solution back to $\mathbb{R}^{3,1}$.
To make this concrete, we start from the Minkowski metric on $\mathbb{R}^{3,1}$,
\begin{eqnarray}
ds^{2} & = & dt^{2}-dx^{2}-dy^{2}-dz^{2}\nonumber \\
& = & d\tau^{2}-dr^{2}-r^{2}d\varphi^{2}-\tau^{2}d\eta^{2}.
\end{eqnarray}
where $\tau$ is the proper time, $\eta$ is the spacetime rapidity, $r$ is the transverse radius, and $\varphi$ is the azimuthal angle. These coordinates are defined as
\begin{eqnarray}
\tau=\sqrt{t^{2}-z^{2}}, &\;& \eta=\frac{1}{2}\ln\left(\frac{t+z}{t-z}\right),\nonumber \\
r=\sqrt{x^{2}+y^{2},} &\;& \varphi=\arctan\left(\frac{y}{x}\right).
\end{eqnarray}

We then perform a Weyl rescaling to map the fluid dynamics from $\mathbb{R}^{3,1}$ to the conformally related spacetime $dS_{3}\times\mathbb{R}$. The rescaled line element $d\hat{s}^{2}$ can be brought into the Gubser coordinates $(\rho,\theta,\varphi,\eta)$ as
\begin{eqnarray}
d\hat{s}^{2} & = & \frac{1}{\tau^{2}}(d\tau^{2}-dr^{2}-r^{2}d\varphi^{2})-d\eta^{2}\nonumber \\
& = & d\rho^{2}-\cosh^{2}\rho(d\theta^{2}+\sin^{2}\theta d\varphi^{2})-d\eta^{2}.\label{eq:Weyl_transform}
\end{eqnarray}
The coordinate transformation is defined by
\begin{eqnarray}
\text{sinh}\rho & = & -\frac{L^{2}-\tau^{2}+r^{2}}{2L\tau},\nonumber \\
\tan\theta & = & \frac{2Lr}{L^{2}+\tau^{2}-r^{2}}.\label{eq:Gubser_coordinate}
\end{eqnarray}
Here $L$ is a parameter with dimensions of length that sets the characteristic transverse size of the system. To satisfy the Gubser symmetry, the velocity of the fluid in the coordinate
$(\rho,\theta,\varphi,\eta)$ is found to be, 
\begin{eqnarray}
\hat{u}_{\mu} & = & (1,0,0,0).\label{eq:_Velocity_of_Gubser_flow}
\end{eqnarray}

According to the Weyl rescaling in Eq.~(\ref{eq:Weyl_transform}), the metric $g_{\mu\nu}$ in Minkowski spacetime and the metric $\hat{g}_{\mu\nu}$ in $dS_{3}\times \mathbb{R}$ are related by
\begin{eqnarray}
    \hat{g}_{\mu\nu}=\frac{1}{\tau^{2}}g_{\mu\nu},\  & \hat{g}^{\mu\nu}=\tau^{2}g^{\mu\nu}.
\end{eqnarray}
The Christoffel symbols transform as
\begin{eqnarray}
\Gamma_{\mu\nu}^{\lambda} & = & \hat{\Gamma}_{\mu\nu}^{\lambda}+\tau^{-1}(\delta_{\nu}^{\lambda}\hat{\nabla}_{\mu}\tau+\delta_{\mu}^{\lambda}\hat{\nabla}_{\nu}\tau-\hat{g}_{\mu\nu}\hat{g}^{\lambda\alpha}\hat{\nabla}_{\alpha}\tau).\label{eq:Transformation_of_Christoffel_symbols}
\end{eqnarray}
The general relation between a physical tensor $A_{\nu_{1}\nu_{2}\cdots\nu_{n}}^{\mu_{1}\mu_{2}\cdots\mu_{m}}(x)$ in $\mathbb{R}^{3,1}$ and its counterpart $\hat{A}_{\nu_{1}\nu_{2}\cdots\nu_{n}}^{\mu_{1}\mu_{2}\cdots\mu_{m}}(x)$ in $dS_{3}\times\mathbb{R}$ is
\begin{eqnarray}
\hat{A}_{\nu_{1}\nu_{2}\cdots\nu_{n}}^{\mu_{1}\mu_{2}\cdots\mu_{m}}(x) & = & \tau^{\Delta_{A}}A_{\nu_{1}\nu_{2}\cdots\nu_{n}}^{\mu_{1}\mu_{2}\cdots\mu_{m}}(x).\label{eq:Weyl_scaling}
\end{eqnarray}
where $\Delta_{A}=[A]+m-n$, and $[A]$ denotes the mass dimension of $A$.

\subsection{Spin hydrodynamics in $dS_3\times \mathbb{R}$ spacetime}
\label{subsec:spin_dS3}

In this subsection, we briefly review the spin hydrodynamic with Gubser flow following our previous studies \cite{Wang:2021wqq}. 
Using this rule in Eq.~\eqref{eq:Weyl_scaling}, it is straightforward to derive that 
\begin{eqnarray}
\hat{\Theta}^{\mu\nu} & = & \tau^{6}\Theta^{\mu\nu}.
\end{eqnarray}
and the energy-momentum conservation equation \eqref{eq:conservation_eq} in 
$dS_3\times \mathbb{R}$ spacetime,
\begin{eqnarray}
0=\nabla_{\mu}\Theta^{\mu\nu} & = & \tau^{-6}(\hat{\nabla}_{\mu}\hat{\Theta}^{\mu\nu}-2\tau^{-1}\hat{\Theta}^{[\mu\nu]}\hat{\nabla}_{\mu}\tau)-\tau^{-7}\hat{\Theta}_{\ \mu}^{\mu}\hat{g}^{\nu\alpha}\hat{\nabla}_{\alpha}\tau.
\end{eqnarray}
Since we set the bulk pressure to vanish, we have $\hat{\Theta}_{\ \mu}^{\mu}=0$. The above equation therefore reduces to
\begin{eqnarray}
\hat{\nabla}_{\mu}\hat{\Theta}^{\mu\nu}-2\tau^{-1}\hat{\Theta}^{[\mu\nu]}\hat{\nabla}_{\mu}\tau & = & 0.\label{eq:Conservation_eq_of_EMT_in_dS_3}
\end{eqnarray}
Similarly, the conservation equation for the total angular momentum tensor in Eq.~(\ref{eq:spin_tensor}) becomes 
\begin{eqnarray}
\hat{\nabla}_{\lambda}(\hat{u}^{\lambda}\hat{S}^{\mu\nu})+2\hat{\phi}^{\mu\nu}-\tau^{-1}\hat{\nabla}_{\sigma}\tau(\hat{g}^{\mu\sigma}\hat{u}_{\alpha}\hat{S}^{\alpha\nu}+\hat{g}^{\nu\sigma}\hat{u}_{\alpha}\hat{S}^{\mu\alpha}+\hat{u}^{\mu}\hat{S}^{\nu\sigma}-\hat{u}^{\nu}\hat{S}^{\mu\sigma}) & = & 0.\label{eq:Spin_evolution_eq_in_dS_3}
\end{eqnarray}
Here, we drop $\Sigma^{\lambda\mu\nu}_{(1)}$ and $q^\mu$ to simplify the subsequent steps in searching for analytic solutions. 
The constitutive equation for $\phi^{\mu\nu}$ in Eq.~\eqref{eq:q_phi} now becomes 
%{\color{blue} Need to be updated!}
%\begin{eqnarray}
%0 & = & \tau_{\phi}\hat{\Delta}^{\mu\alpha}\hat{\Delta}^{\nu\beta}\hat{u}^{\sigma}\hat{\nabla}_{\sigma}\hat{\phi}_{\alpha\beta}-4\tau^{-1}\tau_{\phi}\hat{u}^{\sigma}\hat{\phi}^{\mu\nu}\hat{\nabla}_{\sigma}\tau \nonumber \\
%&&
% -\tau^{4}\gamma_{s}\hat{\Delta}^{\mu\alpha}\hat{\Delta}^{\nu\beta}\left(\hat{\nabla}_{\alpha}\hat{u}_{\beta}-\hat{\nabla}_{\beta}\hat{u}_{\alpha}+4\hat{\omega}_{\alpha\beta}\right)+\tau\hat{\phi}^{\mu\nu},
% \label{eq:Constitutive_eq_in_dS_3}
%\end{eqnarray}
%{\color{red} By Genhui: 
\begin{eqnarray}
0 & = & \tau_{\phi}\hat{\Delta}^{\mu\alpha}\hat{\Delta}^{\nu\beta}\hat{u}^{\sigma}\hat{\nabla}_{\sigma}\hat{\phi}_{\alpha\beta}-4\tau^{-1}\tau_{\phi}\hat{u}^{\sigma}\hat{\phi}^{\mu\nu}\hat{\nabla}_{\sigma}\tau \nonumber \\
&&
 -\tau^{4}\gamma_{s}\hat{\Delta}^{\mu\alpha}\hat{\Delta}^{\nu\beta}\left(\hat{\nabla}_{\alpha}\hat{u}_{\beta}-\hat{\nabla}_{\beta}\hat{u}_{\alpha}+4\hat{\omega}_{\alpha\beta}\right)+\tau\hat{\phi}^{\mu\nu}\nonumber \\
&&
 +
 \lambda_\theta\hat{\phi}^{\mu\nu}\left(\hat{\nabla}_\alpha\hat{u}^\alpha+3\tau^{-1}\hat{u}^{\sigma}\hat{\nabla}_{\sigma}\tau\right),
 \label{eq:Constitutive_eq_in_dS_3}
\end{eqnarray}
%}
where we have used that
\begin{eqnarray}
\nabla_{\mu}u_{\nu} & = & \tau\hat{\nabla}_{\mu}\hat{u}_{\nu}-\hat{u}_{\mu}\hat{\nabla}_{\nu}\tau+\hat{g}_{\mu\nu}\hat{u}^{\alpha}\hat{\nabla}_{\alpha}\tau,\nonumber \\
\nabla_{\sigma}\phi_{\alpha\beta} & = & \tau^{-2}\hat{\nabla}_{\sigma}\hat{\phi}_{\alpha\beta}-\tau^{-3}\left(\hat{\phi}_{\sigma\beta}\hat{\nabla}_{\alpha}\tau+\hat{\phi}_{\alpha\sigma}\hat{\nabla}_{\beta}\tau+4\hat{\phi}_{\alpha\beta}\hat{\nabla}_{\sigma}\tau\right) \nonumber \\& &+\tau^{-3}\left(\hat{g}_{\sigma\alpha}\hat{g}^{\lambda\mu}\hat{\phi}_{\lambda\beta}+\hat{g}_{\sigma\beta}\hat{g}^{\lambda\mu}\hat{\phi}_{\alpha\lambda}\right)\hat{\nabla}_{\mu}\tau.
\end{eqnarray}
From the above equation, we find that conformal symmetry is preserved for the constitutive equation of $\phi^{\mu\nu}$ when 
\begin{eqnarray}
    \lambda_\theta = \frac{4}{3}\tau_\phi.
\end{eqnarray}
For simplicity, we restrict ourselves to this choice in the remainder of this work. In this case, Eq.~\eqref{eq:Constitutive_eq_in_dS_3} reduces to
\begin{eqnarray}
    0&=&\frac{\tau_{\phi}}{\tau}\hat{\Delta}^{\mu\alpha}\hat{\Delta}^{\nu\beta}\hat{u}^{\sigma}\hat{\nabla}_{\sigma}\hat{\phi}_{\alpha\beta}+\frac{4}{3}\frac{\tau_{\phi}}{\tau}\hat{\phi}^{\mu\nu}\hat{\nabla}_{\alpha}\hat{u}^{\alpha}+\hat{\phi}^{\mu\nu}\nonumber\\&&-\tau^{3}\gamma_{s}\hat{\Delta}^{\mu\alpha}\hat{\Delta}^{\nu\beta}\left(\hat{\nabla}_{\alpha}\hat{u}_{\beta}-\hat{\nabla}_{\beta}\hat{u}_{\alpha}+4\hat{\omega}_{\alpha\beta}\right),
    \label{eq:phi_hat}
\end{eqnarray}

To close the system, we consider the equations of state for conformal fluid \cite{Xie:2023gbo,Wang:2024afv},
\begin{eqnarray}
\hat{e} &=& c_s^2 \hat{p},\\
\hat{S}^{\mu\nu}&=&\hat{\chi}\hat{\omega}^{\mu\nu}+\hat{\chi}'\hat{u}_{\lambda}(\hat{\omega}^{\lambda\mu}\hat{u}^{\nu}-\hat{\omega}^{\lambda\nu}\hat{u}^{\mu}). 
\label{eq:EOS}
\end{eqnarray}
where $c_s$ is the speed of sound and $\chi,\chi'$ are two spin susceptibilities satisfying $\chi'>\chi>0$.

Next, we insert the $dS_3\times\mathbb{R}$ metric $\hat{g}_{\mu\nu}=\text{diag}\{1,-\cosh^{2}\rho,-\cosh^{2}\rho\sin^{2}\theta,-1\}$ to derive the differential equations of spin hydrodynamics in $dS_3\times\mathbb{R}$.
Eq.~(\ref{eq:Conservation_eq_of_EMT_in_dS_3})
splits into four distinct component equations, as shown below, 
\begin{eqnarray}
0 & = & \hat{\partial}_{\rho}\hat{e}+2\tanh\rho(\hat{e}+\hat{p}),
\label{eq:energydensity}
\end{eqnarray}
and
\begin{eqnarray}
0 & = & \frac{1}{\cosh^{2}\rho}\hat{\partial}_{\theta}\hat{p}+\hat{\partial}_{\mu}\hat{\phi}^{\mu\theta}-2\tau^{-1}\hat{\phi}^{\mu\theta}\hat{\nabla}_{\mu}\tau,\nonumber \\
0 & = & \frac{1}{\cosh^{2}\rho\sin^{2}\theta}\hat{\partial}_{\varphi}\hat{p}+\hat{\partial}_{\mu}\hat{\phi}^{\mu\varphi}+\frac{1}{\tan\theta}\hat{\phi}^{\theta\varphi}-2\tau^{-1}\hat{\phi}^{\mu\varphi}\hat{\nabla}_{\mu}\tau,\nonumber \\
0 & = & \hat{\partial}_{\eta}\hat{p}+\hat{\partial}_{\mu}\hat{\phi}^{\mu\eta}+\frac{1}{\tan\theta}\hat{\phi}^{\theta\eta}-2\tau^{-1}\hat{\phi}^{\mu\eta}\hat{\nabla}_{\mu}\tau,
\label{eq:conservation_energy_02}
\end{eqnarray}
where we have used $\hat{u}_{\mu}\hat{\phi}^{\mu\nu}=0$ from
$\phi^{\rho\nu}=\phi^{\nu\rho}=0$ and inserted the $\hat{u}^\mu$ in Eq.~\eqref{eq:_Velocity_of_Gubser_flow}.
Eq.~(\ref{eq:Spin_evolution_eq_in_dS_3})
can be expressed as six independent equations for the components of
$S^{\mu\nu}$, 
\begin{eqnarray}
0 & = & \hat{\partial}_{\rho}\hat{S}^{\varphi\eta}+3\tanh\rho\hat{S}^{\varphi\eta}+2\hat{\phi}^{\varphi\eta}.
\label{eq:Srhotheta}
\end{eqnarray}
and
\begin{eqnarray}
0 & = & \hat{\partial}_{\rho}\hat{S}^{\rho\theta}+3\tanh\rho\hat{S}^{\rho\theta},
\nonumber \\
0 & = & \hat{\partial}_{\rho}\hat{S}^{\rho\varphi}+3\tanh\rho\hat{S}^{\rho\varphi}-\tau^{-1}\hat{S}^{\varphi\theta}\hat{\nabla}_{\theta}\tau,\nonumber \\
0 & = & \hat{\partial}_{\rho}\hat{S}^{\rho\eta}+2\tanh\rho\hat{S}^{\rho\eta}-\tau^{-1}\hat{S}^{\eta\theta}\hat{\nabla}_{\theta}\tau,\nonumber \\
0 & = & \hat{\partial}_{\rho}\hat{S}^{\theta\varphi}+4\tanh\rho\hat{S}^{\theta\varphi}+2\hat{\phi}^{\theta\varphi}+\tau^{-1}\frac{1}{\cosh^{2}\rho}\hat{S}^{\rho\varphi}\hat{\nabla}_{\theta}\tau,\nonumber \\
0 & = & \hat{\partial}_{\rho}\hat{S}^{\theta\eta}+3\tanh\rho\hat{S}^{\theta\eta}+2\hat{\phi}^{\theta\eta}+\tau^{-1}\frac{1}{\cosh^{2}\rho}\hat{S}^{\rho\eta}\hat{\nabla}_{\theta}\tau.
\label{eq:S_all}
\end{eqnarray}
We find that the components  $\hat{S}^{\rho\varphi}$, $\hat{S}^{\rho\eta}$, $\hat{S}^{\varphi\theta}$, and $\hat{S}^{\eta\theta}$ are mutually coupled and Eqs.~(\ref{eq:conservation_energy_02}) are not automatically satisfied. Equivalently, the standard Gubser-flow velocity profile in $dS_3\times\mathbb{R}$ spacetime in Eq.~(\ref{eq:_Velocity_of_Gubser_flow}) no longer remains a solution of the dynamical equations during time evolution unless we set  $\hat{S}^{\rho\varphi}$, $\hat{S}^{\rho\eta}$, $\hat{S}^{\varphi\theta}$, and $\hat{S}^{\eta\theta}$ to zero \cite{Wang:2021wqq}.
%Additionally, the solution for $\hat{S}^{\rho\theta}$ is trivial and we will set it to be zero for simplicity.
Interestingly, we find that the evolution equation for $\hat{S}^{\rho\theta}$ decouples from that for $\hat{\phi}^{\rho\theta}$. As a result, the solution for $\hat{S}^{\rho\theta}$ contains no additional information about the relaxation time or dissipative effects. We therefore set the initial condition $\hat{S}^{\rho\theta}(\rho_0)=0$ and do not consider this component in the following.
Therefore, following our previous treatment in Ref.~\cite{Wang:2021wqq}, we retain only a finite $\hat{S}^{\varphi\eta}$ and set all other components of $\hat{S}^{\mu\nu}$ to zero.

With this simplification, the full set of differential equations reduces to three differential equations: Eqs.~(\ref{eq:energydensity}), (\ref{eq:Srhotheta}) and  the constitutive equation for $\hat{\phi}^{\varphi\eta}$ from Eq.~\eqref{eq:phi_hat}, %{\color{blue} Need to be updated!}
%\begin{eqnarray}
%0 & = & \hat{\partial}_{\rho}\hat{\phi}^{\varphi\eta}+\left(\tanh\rho+\frac{\tau}{\tau_{\phi}}-4\tau^{-1}\hat{\nabla}_{\rho}\tau\right)\hat{\phi}^{\varphi\eta}-\frac{4\tau^{4}\gamma_{s}}{\hat{\chi}\tau_{\phi}}\hat{S}^{\varphi\eta}.\label{eq:phiphieta}
%\end{eqnarray}
%{\color{red} By Genhui:
\begin{eqnarray}
0 & = & \hat{\partial}_{\rho}\hat{\phi}^{\varphi\eta}+\left(\frac{11}{3}\tanh\rho+\frac{\tau}{\tau_{\phi}}\right)\hat{\phi}^{\varphi\eta}-\frac{4\tau^{4}\gamma_{s}}{\hat{\chi}\tau_{\phi}}\hat{S}^{\varphi\eta}.\label{eq:phiphieta}
\end{eqnarray}
%}
Eq.~(\ref{eq:energydensity}) gives 
exactly the same analytic solution for $\hat{e}$ as in the standard Gubser flow.
We combine Eqs.~(\ref{eq:Srhotheta}) and (\ref{eq:phiphieta}) and derive the equation for 
 $\hat{S}^{\varphi\eta}$: %becomes 
\begin{equation}
0=\frac{d^{2}}{d\rho^{2}}\hat{S}+\left(\frac{20}{3}\tanh\rho+\hat{\tau}_{\phi}^{-1}\right)\frac{d}{d\rho}\hat{S}\\+\left(3+8\tanh^{2}\rho+3\hat{\tau}_{\phi}^{-1}\tanh\rho+8\hat{\gamma}_{s}\hat{\tau}_{\phi}^{-1}\hat{\chi}^{-1}\right)\hat{S},
\label{eq:Eq_S}
\end{equation}
where  
\begin{equation}
\hat{\tau}_{\phi}=\frac{\tau_{\phi}}{\tau},\quad\hat{\gamma}_{s}=\tau^{3}\gamma_{s},\label{eq:tau_phi_hat}
\end{equation}
are the Weyl-rescaled relaxation time associated with $\phi^{\mu\nu}$ and the Weyl-rescaled transport coefficient $\gamma_s$, respectively. After the Weyl transformation, we assume that $\hat{\tau}_{\phi}$ and $\hat{\gamma}_{s}$ depend on $\rho$ only. From now on, we denote $\hat{S}^{\varphi\eta}$ by $\hat{S}$. It is then clear that $\hat{S}$ is a function of $\rho$ only, consistent with our Gubser-flow assumption.

\section{Asymptotic solutions at late time}
\label{sec:Asymptotic_solutions}

In the previous section, we derived the main differential equation for $\hat{S}$, given in Eq.~(\ref{eq:Eq_S}). We now study the asymptotic behavior of $\hat{S}$ at late times, i.e., in the limit $\rho\to +\infty$. Since Eq.~(\ref{eq:Eq_S}) contains several factors of $\tanh\rho$, which approach unity as $\rho\to\infty$, it is convenient to introduce the variable
\begin{equation}
w=\frac{1}{1-\tanh\rho},
\end{equation}
so that $w\to\infty$ as $\rho\to\infty$.
Let us discuss the physical meaning of $w$. For this purpose, we recall Eq.~(\ref{eq:Gubser_coordinate}), which yields
\begin{equation}
\frac{\tau}{L}=-\frac{1}{(\text{sinh}\rho-\text{cos}\theta\text{cosh}\rho)}.
\end{equation}
At fixed $\theta$, one finds
\begin{equation}
\frac{\tau}{L}=\sqrt{2w-1},\quad\textrm{when }\cos\theta=1.\label{eq:tau_w}
\end{equation}
Therefore, the large $w$ limit corresponds to late proper time limit, i.e., $\tau^{2}\rightarrow\infty$ at fixed $\cos\theta$. Next, we introduce an auxiliary function $f$ defined by
\begin{eqnarray}
f(w) & = & \frac{(2w-1)}{\hat{S}}\frac{d\hat{S}}{dw}. 
\label{eq:f_S}
\end{eqnarray}
Once $f(w)$ is determined, $\hat{S}$ can be reconstructed from $f(w)$ via
\begin{eqnarray}
\text{ln}\left[\frac{\hat{S}(w)}{\hat{S}(w_{0})}\right] & = & \int_{w_{0}}^{w}\frac{f(w^{\prime})}{2w^{\prime}-1}dw^{\prime}.
\end{eqnarray}

To study the dynamical evolution of $f(w)$, we follow Refs.~\cite{Blaizot:2021cdv,Wang:2024afv} and parameterize the transport coefficients as
\begin{eqnarray}
\hat{\tau}_{\phi}^{-1} & = & \alpha w^{\Delta_{1}},\nonumber \\
8\hat{\gamma}_{s}\hat{\tau}_{\phi}^{-1}\hat{\chi}^{-1} & = & \beta w^{\Delta_{2}},
\label{eq:Delta_12}
\end{eqnarray}
where $\alpha$, $\beta$, and $\Delta_{1,2}$ are constants. Since the transport coefficients $\hat{\gamma}_{s}$ and $\hat{\tau}_{\phi}^{-1}$, as well as the spin susceptibility $\hat{\chi}$, are positive-definite as required by the entropy principle (see, e.g., the discussion below Eqs.~(\ref{eq:q_phi}, \ref{eq:EOS}) and the references therein), the constants $\alpha$ and $\beta$ must also be positive-definite. 
On the other hand, according to Eqs.~(\ref{eq:tau_phi_hat}, \ref{eq:tau_w}), one finds $\tau_{\phi}\sim\tau^{1-2\Delta_{1}}$. Returning to the constitutive equation for $\phi^{\mu\nu}$ in Eq.~(\ref{eq:con_phi}), the relaxation-time term $\tau_{\phi}\Delta_{\alpha}^{\mu}\Delta_{\beta}^{\nu}u^{\rho}\nabla_{\rho}\phi^{\alpha\beta}\sim\tau_{\phi}\frac{d}{d\tau}\phi^{\mu\nu}$ should, in general, be of higher order than $\phi^{\mu\nu}$ in the hydrodynamic gradient expansion, i.e., $\left|\tau_{\phi}\frac{d}{d\tau}\phi^{\mu\nu}\right|/\left|\phi^{\mu\nu}\right|\ll1$.
This implies
\begin{equation}
\Delta_{1}>0.
\end{equation}
By contrast, there is no strict constraint on $\Delta_{2}$.
After substituting into Eq.~(\ref{eq:Eq_S}), we obtain
\begin{eqnarray}
0 & = & w^{2}\frac{df}{dw}+\frac{w^{2}}{(2w-1)}f^{2}+\frac{w}{(2w-1)}\left[\frac{20}{3}(w-1)+\alpha w^{\Delta_{1}+1}\right]f\nonumber \\
 &  & +\frac{w}{(2w-1)}\left(11w-16+8w^{-1}+3\alpha w^{\Delta_{1}+1}-3\alpha w^{\Delta_{1}}+\beta w^{\Delta_{2}+1}\right).\label{eq:f_w_02}
\end{eqnarray}

%\subsection{Asymptotic solutions for $f(w)$ at late time}

Next, we implement the slow-roll expansion method \cite{Strickland:2017kux,Denicol:2018pak} to derive the asymptotic solutions. For convenience, we redefine $f(w)$ in terms of a new function $v(w)$ as
\begin{equation}
v(w)=\frac{1}{2w-1}f(w)+\frac{1}{2}\mathcal{A}(w),\label{eq:def_v}
\end{equation}
where 
\begin{equation}
\mathcal{A}(w)=\frac{26-20w^{-1}+3\alpha w^{\Delta_{1}}}{3(2w-1)}.
\end{equation}
With this change of variables, Eq.~(\ref{eq:f_w_02}) reduces to the Riccati-type equation
\begin{equation}
\frac{dv(w)}{dw}+v(w)^{2}=\mathcal{I}(w),\label{eq:diff_v}
\end{equation}
where the source term reads
\begin{eqnarray}
\mathcal{I}(w) & = & \frac{1}{(2w-1)^{2}}\left[-\frac{8}{9}-\frac{2}{9}w^{-2}+\frac{4}{9}w^{-1}+\left(\frac{1}{3}+\Delta_{1}\right)\alpha w^{\Delta_{1}}\right.\nonumber \\
 &  & \left.-\left(\frac{1}{2}\Delta_1+\frac{1}{3}\right)\alpha w^{\Delta_{1}-1}+\frac{1}{4}\alpha^{2}w^{2\Delta_{1}}-\beta w^{\Delta_{2}}\right].\label{eq:I_01}
\end{eqnarray}
We now note that Eq.~(\ref{eq:diff_v}) contains no term linear in $v(w)$. Next, we introduce an auxiliary parameter $\epsilon\ll1$. We emphasize that $\epsilon$ is merely a book-keeping parameter, which will be set to $1$ at the end. We then expand $v(w)$ in a power series in $\epsilon$,
\begin{equation}
v(w)=v_{0}(w)+\epsilon v_{1}(w)+\epsilon^{2}v_{2}(w)+\mathcal{O}(\epsilon^{3}).\label{eq:v_expansion}
\end{equation}

\begin{table}[t]
\centering

\caption{Asymptotic solutions for $f(w)$ in the large $w$ limit for different
choices of $\Delta_{1,2}$. Here, $k_{1,2}$ and $\mathcal{I}_{1,2}$
are defined in Eq.~(\ref{eq:I_02}). }
\label{tab:Solutions_f}

\centering{}%
\begin{tabular}{c|c|c|c|c|c}
\hline 
 & $\mathcal{I}_{1}$  & $k_{1}$  & $\mathcal{I}_{2}$  & $k_{2}$  & $f_{\pm}(w)$\tabularnewline
\hline 
\hline 
$\Delta_{1}>\Delta_{2}$  & $\frac{1}{16}\alpha^{2}$  & $2\Delta_{1}-2$  & $\frac{1}{4}\left(\frac{1}{3}+\Delta_{1}\right)\alpha$  & $\Delta_{1}-2$  & $\begin{cases}
f_{+}= & -3\\
f_{-}= & -\alpha w^{\Delta_{1}}
\end{cases}$\tabularnewline
\hline 
$\Delta_{1}=\Delta_{2}$  & $\frac{1}{16}\alpha^{2}$  & $2\Delta_{1}-2$  & $\frac{1}{4}\left[\left(\frac{1}{3}+\Delta_{1}\right)\alpha-\beta\right]$  & $\Delta_{1}-2$  & $\begin{cases}
f_{+}= & -\left(3+\frac{\beta}{\alpha}\right)\\
f_{-}= & -\alpha w^{\Delta_{1}}
\end{cases}$\tabularnewline
\hline 
$\Delta_{1}<\Delta_{2}<2\Delta_{1}$  & $\frac{1}{16}\alpha^{2}$  & $2\Delta_{1}-2$  & $-\frac{\beta}{4}$  & $\Delta_{2}-2$  & $\begin{cases}
f_{+}= & -\frac{\beta}{\alpha}w^{\Delta_{2}-\Delta_{1}}\\
f_{-}= & -\alpha w^{\Delta_{1}}
\end{cases}$\tabularnewline
\hline 
$\Delta_{1}<\Delta_{2}=2\Delta_{1}$  & $\frac{\alpha^{2}-4\beta}{16}$  & $2\Delta_{1}-2$  & $\frac{1}{4}\left(\frac{1}{3}+\Delta_{1}\right)\alpha$  & $\Delta_{1}-2$  & $f_{\pm}=\frac{1}{2}\left(-\alpha\pm\sqrt{\alpha^{2}-4\beta}\right)w^{\Delta_{1}}$\tabularnewline
\hline 
$\Delta_{1}<2\Delta_{1}<\Delta_{2}$  & $-\frac{\beta}{4}$  & $\Delta_{2}-2$  & $\frac{1}{16}\alpha^{2}$  & $2\Delta_{1}-2$  & $f_{\pm}=\pm i\sqrt{\beta}w^{\frac{\Delta_{2}}{2}}-\frac{\alpha}{2}w^{\Delta_{1}}$\tabularnewline
\hline 
\end{tabular}
\end{table}

In the slow roll expansion, each derivative is counted as an additional power of $\epsilon$. Accordingly, Eq.~(\ref{eq:diff_v}) is rewritten as 
\begin{equation}
\epsilon\frac{dv(w)}{dw}+v(w)^{2}=\mathcal{I}(w).
\end{equation}
Substituting Eq.~(\ref{eq:v_expansion}) into the above equation, we obtain
\begin{equation}
\left[v_{0}^{2}-\mathcal{I}(w)\right]+\epsilon\left[\frac{dv_{0}}{dw}+2v_{0}v_{1}\right]+\epsilon^{2}\left[\frac{dv_{1}}{dw}+v_{1}^{2}+2v_{0}v_{2}\right]+\mathcal{O}(\epsilon^{3})=0.
\end{equation}
Since $\epsilon$ is arbitrary, the above equation can be satisfied order by order only if the coefficients of each power $\epsilon^{n}$ vanish. This yields
\begin{eqnarray}
v_{0}^{2}-\mathcal{I}(w) & = & 0,\nonumber \\
\frac{dv_{0}}{dw}+2v_{0}v_{1} & = & 0,\nonumber \\
\frac{dv_{1}}{dw}+v_{1}^{2}+2v_{0}v_{2} & = & 0.
\end{eqnarray}
These equations can be solved sequentially. The resulting slow-roll expansion for $v$ is
\begin{equation}
v(w)=\pm\sqrt{\mathcal{I}}-\frac{\epsilon}{4}\frac{d\ln\mathcal{I}}{dw}\mp\frac{\epsilon^{2}}{32\mathcal{I}^{\frac{5}{2}}}\left[-4\mathcal{I}\frac{d^{2}\mathcal{I}}{dw^{2}}+5\left(\frac{d\mathcal{I}}{dw}\right)^{2}\right]+\mathcal{O}(\epsilon^{3}).\label{eq:v_02}
\end{equation}
Since we are interested in the large $w$ behavior, we further simplify $\mathcal{I}(w)$ in the $w\to\infty$ limit. In general, we may expand $\mathcal{I}(w)$ in Eq.~(\ref{eq:I_01}) as
\begin{equation}
\mathcal{I}(w)=\mathcal{I}_{1}w^{k_{1}}+\mathcal{I}_{2}w^{k_{2}}+...,\label{eq:I_02}
\end{equation}
where $k_{1}>k_{2}$. We note that, given the explicit form of $\mathcal{I}(w)$ in Eq.~\eqref{eq:I_01}, the leading power in the large $w$ expansion satisfies $k_{1}>-2$. 
Here, $\mathcal{I}_{1}w^{k_{1}}$ and $\mathcal{I}_{2}w^{k_{2}}$ denote the leading and next-to-leading terms, respectively, in the large $w$ expansion. The exponents $k_{1,2}$ depend on the specific values of $\Delta_{1,2}$.
Setting $\epsilon\rightarrow1$ and substituting Eq. (\ref{eq:I_02}) into Eq.~ (\ref{eq:v_02}), we obtain
\begin{equation}
v(w)=\pm\sqrt{\mathcal{I}_{1}}w^{\frac{k_{1}}{2}}\pm\frac{\mathcal{I}_{2}}{2\sqrt{\mathcal{I}_{1}}}w^{k_{2}-\frac{k_{1}}{2}}-\frac{k_{1}}{4}w^{-1}+...,
\end{equation}
where $...$ denotes terms that are high order terms in the large $w$ expansion.

After some algebra, and using Eq.~(\ref{eq:def_v}), we derive the corresponding asymptotic solutions for $f(w)$ for different choices of $\Delta_{1,2}$, which are summarized in Table~\ref{tab:Solutions_f}. Let us briefly discuss the results summarized in Table~\ref{tab:Solutions_f}. Recalling Eq.~\eqref{eq:f_S}, the late time behavior of $\hat{S}$ can be obtained by integrating the corresponding branches $f_{\pm}(w)$. In particular, complex (imaginary) asymptotic forms of $f_{\pm}(w)$ imply oscillatory behavior of $\hat{S}$. For this reason, we keep the complex solutions for $f(w)$ in Table~\ref{tab:Solutions_f}.

We can also cross-check the above results by a different method. As an example, consider the case $\Delta_{1}>\Delta_{2}$ with $\Delta_{1}>0$. In this regime, keeping only the leading contributions to Eq.~(\ref{eq:f_w_02}) in the large $w$ limit yields
\begin{equation}
2w^{2}\frac{df}{dw}+wf^{2}+\alpha w^{\Delta_{1}+1}f+3\alpha w^{\Delta_{1}+1} =0.\label{eq:temp_f_01}
\end{equation}
Under these assumptions, there are two consistent dominant balances in the large $w$ expansion. First, if the leading terms are $wf^{2}+\alpha w^{\Delta_{1}+1}f$, one obtains $f_{-}\rightarrow-\alpha w^{\Delta_{1}}$. Second, if the last two terms $\alpha w^{\Delta_{1}+1}f+3\alpha w^{\Delta_{1}+1}$ dominate, one finds $f_{+}\rightarrow-3$.
We further assess the stability of the two asymptotic solutions $f_{\pm}$. Writing $f\rightarrow f_{\pm}+\delta f_{\pm}$ and substituting into Eq. (\ref{eq:temp_f_01}), we obtain $\delta f_{\pm}\sim e^{\mp\frac{1}{2\Delta_{1}}\alpha w^{\Delta_{1}}}$. Hence $\delta f_{+}$ decreases with $w$, whereas $\delta f_{-}$ grows with $w$. In other words, $f_{+}$ is stable and $f_{-}$ is unstable.  
In the next section, we will show that the stable asymptotic solutions correspond to possible attractors, whereas the unstable asymptotic solutions correspond to repulsive behavior (i.e., repellers).
The analysis can be extended straightforwardly to other parameter choices.

\section{Late time attractor}
\label{sec:attractor}

%1. Put the figures here. Explain the red and blue lines which stand for the attractor and repeller. 

%1(b) Explain why there must exist both attractor and repeller from mathematica.

%1(c) Mention that attractor is always larger than the repeller. 

%2. Explain (a)(b) correspond to the constant aymptotic solution, while (c)(d) correspond to the decay of $f$. Make a connection to the next section. 

%3. Explain why $f\rightarrow -\infty$ and it connects to the $\hat{S}\rightarrow0$ limit. 

%3(a) basin of attraction. The system’s state will eventually converge to the fixed point, provided the initial condition lies within its basin of attraction.

\begin{figure}[tph]
    \centering
    \includegraphics[width=0.8\linewidth]{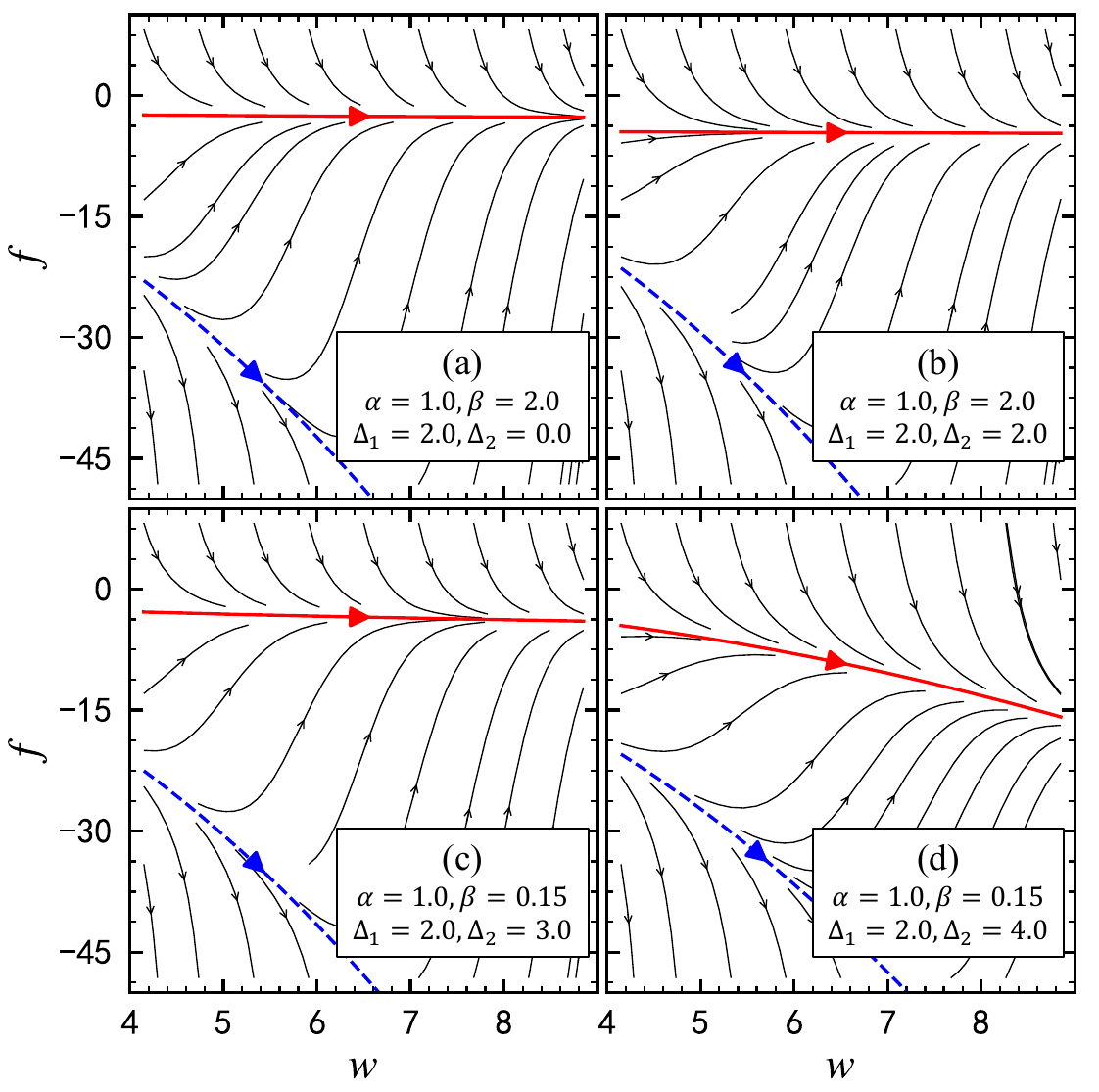}
    \caption{Asymptotic behavior of $f(w)$ at large $w$ for different parameter choices. The red solid and blue dashed curves denote the attractors and repellers summarized in Table~\ref{tab:Solutions_f}, respectively. The gray solid curves show numerical solutions for different initial conditions. }
    \label{fig:attractor}
\end{figure}

We show the large $w$ asymptotic behavior of $f(w)$ for different choices of $\alpha$, $\beta$, $\Delta_1$, and $\Delta_2$ in Fig.~\ref{fig:attractor}. The red solid and blue dashed curves correspond to the asymptotic solutions summarized in Table~\ref{tab:Solutions_f}, while the gray solid curves represent numerical solutions obtained from different initial conditions. We observe that, for a wide range of initial conditions, the numerical solutions approach the red curves and move away from the blue dashed ones. We therefore identify the red solid and blue dashed curves as the attractors and repellers, respectively.

Panels (a) and (b) correspond to the two cases with constant asymptotic solutions, namely $\Delta_{1}>\Delta_{2}$ with $f(w)\rightarrow-3$, and $\Delta_{1}=\Delta_{2}$ with $f(w)\rightarrow-\left(3+\frac{\beta}{\alpha}\right)$. Panels (c) and (d) correspond to cases in which $f(w)$ exhibits a power-law decaying behavior in the large $w$ limit, namely $\Delta_{1}<\Delta_{2}<2\Delta_{1}$ with $f_{+}=-\frac{\beta}{\alpha}w^{\Delta_{2}-\Delta_{1}}$, and $\Delta_{1}<\Delta_{2}=2\Delta_{1}$ with $f_{+}=\frac{1}{2}\left(-\alpha+\sqrt{\alpha^{2}-4\beta}\right)w^{\Delta_{1}}$. The numerical solutions are consistent with the asymptotic analysis in the previous section. In the next section, we transform $f(w)$ back to $\hat{S}$ and to the spin density in conventional $(t,x,y,z)$ coordinates. In particular, the cases shown in panels (a) and (b) lead to a power-law decay of $\hat{S}$ at late times.

From Fig.~\ref{fig:attractor}, we observe that, for each fixed set of parameters $(\alpha,\beta,\Delta_{1},\Delta_{2})$, there exists a unique attractor and a unique repeller in the large $w$ regime. Moreover, in the large $w$ limit, the attractor branch takes larger values of $f(w)$ than the repeller branch for the parameter sets shown.
It can be understood from the structure of our evolution equation for $\hat{S}$ in Eq.~(\ref{eq:Eq_S}). As shown in the previous section, after an appropriate change of variables, the differential equation~(\ref{eq:Eq_S}) can be cast into the Riccati-type form given in Eq.~(\ref{eq:diff_v}).

We can perform an additional transformation of $v(w)$ in Eq.~(\ref{eq:diff_v}) by introducing
\begin{equation}
v(w)=\frac{\mathcal{V}^{\prime}(w)}{\mathcal{V}(w)},
\end{equation}
where the prime denotes differentiation with respect to $w$, i.e., $\mathcal{V}^{\prime}(w)=d\mathcal{V}/dw$. Since Eq.~(\ref{eq:diff_v}) contains no term linear in $v(w)$, it can be mapped to the second-order linear equation
\begin{equation}
\mathcal{V}^{\prime\prime}(w)-\mathcal{I}(w)\mathcal{V}(w)=0.\label{eq:V_01}
\end{equation}
In the large $w$ limit, the asymptotic solutions of Eq.~(\ref{eq:V_01}) can be written formally as
\begin{equation}
\mathcal{V}(w)=\exp\left[\pm I_0(w)\right],\quad I_{0}(w)=\int_{w_0}^{w} du\sqrt{\mathcal{I}_{0}(u)},
\end{equation}
where $\mathcal{I}_{0}(w)$ is the leading order of $\mathcal{I}(w)$ as $w\rightarrow\infty$. Note that $\mathcal{I}_0^{\prime\prime}(w)$ does not contribute at leading order in the large-$w$ expansion.
%$\mathcal{I}_{0}(w)=\lim_{w\rightarrow\infty}\mathcal{I}(w)$. 
For real $f(w)$ shown in Fig.~\ref{fig:attractor},  b$\mathcal{I}_{0}(w)$ should be strictly positive. Furthermore, Eq.~\eqref{eq:I_01} implies $I_{0}(w)\sim w^{\delta}$ with $\delta>0$ since $\mathcal{I}_{0}(w)\gg w^{-2}$ as $w\rightarrow\infty$. The general solution for $\mathcal{V}(w)$ is then
\begin{equation}
\mathcal{V}(w)=c_{1}e^{I_{0}(w)}+c_{2}e^{-I_{0}(w)},
\end{equation}
where $c_{1,2}$ are constants determined by the initial conditions. Accordingly, $v(w)$ can be written as
\begin{equation}
v(w)=\sqrt{\mathcal{I}_{0}(w)}\frac{c_{1}-c_{2}e^{-2I_{0}(w)}}{c_{1}+c_{2}e^{-2I_{0}(w)}}.
\end{equation}
If $c_{1}\neq0$, then $e^{-2I_{0}(w)}$ decays rapidly at large $w$, and therefore $v(w)\rightarrow\sqrt{\mathcal{I}_{0}(w)}$. This implies that, for generic initial conditions (i.e., $c_{1}\neq0$), all trajectories approach the branch $v(w)\rightarrow\sqrt{\mathcal{I}_{0}(w)}$ at late times, identifying it as the attractor. In contrast, if $c_{1}=0$, one has $v(w)=-\sqrt{\mathcal{I}_{0}(w)}$. A small perturbation away from this branch generates a small but finite $c_{1}$, after which the trajectory departs from $v(w)=-\sqrt{\mathcal{I}_{0}(w)}$ and approaches the attractor $v(w)\rightarrow\sqrt{\mathcal{I}_{0}(w)}$. This branch therefore corresponds to the repeller.
This analysis shows that there exists a unique attractor and a unique repeller in the large $w$ regime.

In Fig.~\ref{fig:attractor}, we also observe that, in each panel, some gray trajectories lie below the blue dashed curve (the repeller) and keep decreasing toward $-\infty$. We emphasize that this behavior does not indicate a physical singularity in $\hat{S}$. Recalling the definition of $f(w)$ in Eq.~(\ref{eq:f_S}), one can see that $f(w)$ necessarily passes through $\pm\infty$ when $\hat{S}\rightarrow0$. Therefore, trajectories of $f(w)$ that run to $-\infty$ simply correspond to $\hat{S}$ crossing zero: once $\hat{S}$ changes sign, the ratio defining $f(w)$ becomes positive again, and the solution for $f(w)$ re-enters from $+\infty$ (i.e., $f(w)$ effectively flips sign by passing through $\pm\infty$) and subsequently approaches the red solid curve (the attractor). A similar discussion can be found in our previous study~\cite{Wang:2024afv} of attractors in spin hydrodynamics with Bjorken flow.

Based on the discussion above, we can characterize the basin of attraction. For generic initial conditions, the system converges to the attractor at late times; the only exception is the measure-zero set of initial conditions that lie exactly on the repeller branch.

\begin{figure}[t]
    \centering
    \includegraphics[width=0.66\linewidth]{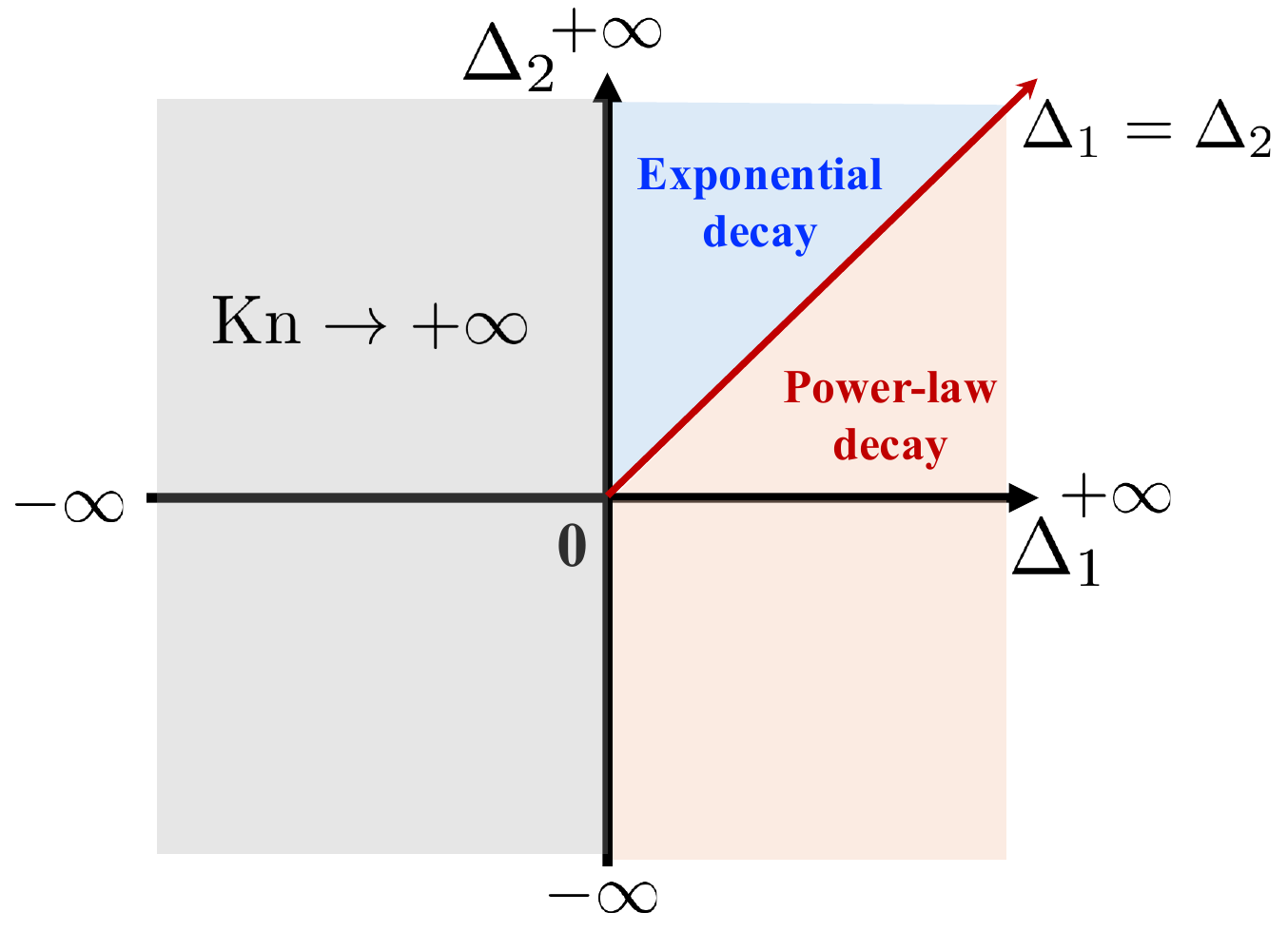}
    \caption{Late-time asymptotic behavior of $\hat{S}$ for different choices of $\Delta_{1,2}$. Here, we take $\tau/\tau_\phi$ as an effective Knudsen number, denoted by $\mathrm{Kn}$. }
    \label{fig:region}
\end{figure}

\section{Late-time power-law decay of spin density} 
\label{sec:S_late_time}

In the conventional relativistic hydrodynamics, the typical decay behavior of the charge number density $n$ is $n\sim\tau^{-1}$ in a Bjorken flow. Therefore, if the spin density exhibits a power-law decay, it can remain comparable at late time. By contrast, if the spin density exhibits an exponential decay, it can be neglected at late time in comparison with the charge number density. Therefore, in this section, we discuss the power-law decay of $\hat{S}$ at late time and the physical picture behind it.

Eq.~\eqref{eq:f_S} shows that if $f(w)\sim-w^{m}$ with $m\neq0$,
then $\hat{S}\sim e^{-w^{m}}$ decays exponentially in $w$, which
is much faster than the typical power-law dilution of a number density.
In such cases, any accumulation effect in $\hat{S}$ would be difficult
to observe. By contrast, when $f(w)\sim\text{const.}$, $\hat{S}$
exhibits a power-law behavior at late times. We find only two possible
late time power-law behaviors for $\hat{S}$ according to the Table~\ref{tab:Solutions_f},
\begin{eqnarray}
(i) \; & \Delta_{1}>\Delta_{2}, & f_+(w)\rightarrow-3,\;\hat{S}\propto w^{-3/2},\nonumber \\
(ii)\; & \Delta_{1}=\Delta_{2}, & f_+(w)\rightarrow-\left(3+\frac{\beta}{\alpha}\right),\;\hat{S}\propto w^{-(3\alpha+\beta)/(2\alpha)}.
\end{eqnarray}
We also summarize it in Fig.~\ref{fig:region}.

In case (i), using Eq.~(\ref{eq:Weyl_scaling}), we obtain 
\begin{equation}
S^{\varphi\eta}=\tau^{-5}\hat{S}^{\varphi\eta}\propto\frac{1}{\tau^{2}}\frac{L^{3}}{\left(L^{2}-\tau^{2}+r^{2}\right)^{3}}.\label{eq:S_late_time_01}
\end{equation}
We note that there are two distinct limits in which one can extract the asymptotic behavior of $S^{\varphi\eta}$ at late proper time $\tau$. If $L\gg\tau$, Eq.~(\ref{eq:S_late_time_01}) reduces to
\begin{equation}
S^{\varphi\eta}\sim\frac{1}{L^{3}\tau^{2}}.
\end{equation}
We can further transform back to the $(t,x,y,z)$ coordinates in flat spacetime and find that the non-vanishing components of the spin density behave as
\begin{equation}
S^{tx},S^{ty},S^{zx},S^{zy}\sim\frac{1}{\tau L^{2}},
\end{equation}
which is consistent with our previous results in Ref.~\cite{Wang:2021wqq}.
In terms of the proper-time scaling, this power-law decay is comparable to that of the conventional charge number density in Bjorken flow. On the other hand, if $\tau\gg L$, Eq.~(\ref{eq:S_late_time_01}) gives
\begin{equation}
S^{\varphi\eta}\sim\frac{L^{3}}{\tau^{8}},
\end{equation}
which corresponds to the non-vanishing components
\begin{equation}
S^{tx},S^{ty},S^{zx},S^{zy}\sim\frac{L^{4}}{\tau^{7}}.
\end{equation}
Although the spin density in this limit still exhibits a power-law decay, it decreases much more rapidly as a function of $\tau$ than other macroscopic hydrodynamic variables. Consequently, its contribution is expected to be strongly suppressed at late proper time.

In case (ii), following a similar analysis, we obtain in the $L\gg\tau$ limit
\begin{equation}
S^{\varphi\eta}\sim\frac{\tau^{(\beta-2\alpha)/\alpha}}{L^{(\beta+3\alpha)/\alpha}},
\label{eq:S_case_ii}
\end{equation}
which implies that the non-vanishing components of the spin density scale as
\begin{equation}
S^{tx},S^{ty},S^{zx},S^{zy}\sim\frac{\tau^{(\beta-\alpha)/\alpha}}{L^{(\beta+2\alpha)/\alpha}}.
\label{eq:S_beta_alpha_01}
\end{equation}
In the opposite limit $\tau\gg L$, we find
\begin{eqnarray}
    S^{\varphi\eta}\sim\frac{L^{(\beta+3\alpha)/\alpha}}{\tau^{(\beta+8\alpha)/\alpha}},
\end{eqnarray}
which gives the non-vanishing components of the spin density as
\begin{equation}
S^{tx},S^{ty},S^{zx},S^{zy}\sim\frac{L^{(\beta+4\alpha)/\alpha}}{\tau^{(\beta+7\alpha)/\alpha}}.
\end{equation}
Interestingly, Eq.~\eqref{eq:S_beta_alpha_01} shows that the spin density scales with proper time as $\tau^{(\beta-\alpha)/\alpha}$.
As mentioned around Eq.~(\ref{eq:Delta_12}), the physical constraint only requires that $\alpha$ and $\beta$ are positive. We note that when $\beta=\alpha$ the spin density in Eq.~\eqref{eq:S_beta_alpha_01} is proper time independent, whereas for $\beta>\alpha$ it can even increase with proper time. Here we emphasize that Eq.~(\ref{eq:S_beta_alpha_01}) is derived in the $L\gg\tau$ limit. Interpreting $L$ as the characteristic size of the fireball produced in the collision, this regime may not be realized in phenomenologically relevant situations.

To explain the physics in cases (i) and (ii), we return to Eqs.~(\ref{eq:Srhotheta}, \ref{eq:phiphieta}). Eq.~(\ref{eq:phiphieta}) follows from Eq.~(\ref{eq:phi_hat}). Using Eqs.~(\ref{eq:Delta_12}) and $\hat{\tau}_{\phi}\sim\alpha^{-1}w^{-\Delta_{1}}$, 
if $\left|\hat{\tau}_{\phi}\hat{\Delta}^{\mu\alpha}\hat{\Delta}^{\nu\beta}\hat{u}^{\sigma}\hat{\nabla}_{\sigma}\hat{\phi}_{\alpha\beta}\right|,\left|\frac{4}{3}\frac{\tau_{\phi}}{\tau}\hat{\phi}^{\mu\nu}\hat{\nabla}_{\alpha}\hat{u}^{\alpha}\right|\ll\left|\hat{\phi}^{\mu\nu}\right|$,
then Eq.~(\ref{eq:phi_hat}) reduces to the algebraic relation $\hat{\phi}^{\mu\nu}\simeq-4\tau^{3}\gamma_{s}\hat{\Delta}^{\mu\alpha}\hat{\Delta}^{\nu\beta}\hat{\omega}_{\alpha\beta}$. This corresponds to a regime in which the effective relaxation time $\hat{\tau}_{\phi}$ decreases more rapidly than the characteristic evolution time of $\hat{\phi}^{\mu\nu}$, so that the dissipative term proportional to $\gamma_{s}$ dominates the evolution of $\hat{\phi}^{\mu\nu}$.
This implies that, when the perturbation is small and driven by dissipative effects, the effective response time $\hat{\tau}_{\phi}$ is too short for $\hat{\phi}^{\mu\nu}$ to develop through its own relaxation dynamics. Instead, $\hat{\phi}^{\mu\nu}$ rapidly approaches an algebraic form determined by the dissipative source term. We then insert this relation into the evolution equation for $\hat{S}^{\varphi\eta}$ in Eq.~(\ref{eq:Srhotheta}).
Compared with Eq.~\eqref{eq:spin_tensor}, Eq.~(\ref{eq:Srhotheta}) can be interpreted as a local balance equation implied by total angular-momentum conservation. The first two terms, $\hat{\partial}_{\rho}\hat{S}^{\varphi\eta}+3\tanh\rho,\hat{S}^{\varphi\eta}$, describe the evolution of the spin tensor, while the source term $-2\hat{\phi}^{\varphi\eta}$ represents the local exchange with orbital angular momentum, i.e., the spin--orbit transfer that redistributes angular momentum between spin and orbital sectors while preserving the total.
Next, we compare these two types of variations. From the scaling of the source term, one finds $-2\hat{\phi}^{\varphi\eta}\sim\frac{\beta}{\alpha}w^{\Delta_{2}-\Delta_{1}}\hat{S}^{\varphi\eta}$. If $\Delta_{1}>\Delta_{2}$, i.e. in case (i), the exchange with orbital angular momentum is parametrically faster than the spin-tensor evolution and dominates over both $\hat{\partial}_{\rho}\hat{S}^{\varphi\eta}$ and the expansion term $3\tanh\rho\,\hat{S}^{\varphi\eta}\sim\hat{S}^{\varphi\eta}(\hat{\nabla}\cdot\hat{u})$. At late times, the remaining evolution of the spin density is then primarily controlled by the expansion term, $3\tanh\rho\,\hat{S}^{\varphi\eta}$, which leads to the power-law decay in Eq.~\eqref{eq:S_late_time_01}.

If $\Delta_{1}=\Delta_{2}$, i.e. in case (ii), the exchange with orbital angular momentum remains comparable to the spin-tensor evolution at late times. Combining the source term with the expansion term then yields $\hat{\partial}_{\rho}\hat{S}^{\varphi\eta}\sim-(3+\frac{\beta}{\alpha})\hat{S}^{\varphi\eta}$ at late times, which gives the power-law behavior in Eq.~(\ref{eq:S_case_ii}).
We then briefly discuss the nontrivial behavior shown in Eq.~(\ref{eq:S_beta_alpha_01}).
Recalling Eqs.~(\ref{eq:tau_phi_hat}, \ref{eq:Delta_12}), we find that, when $\Delta_{1}=\Delta_{2}$, $\beta/\alpha=8\hat{\gamma}_{s}/\hat{\chi}=8\gamma_{s}\tau/\chi$
where $\chi$ is the spin susceptibility in flat Minkowski space. Since $\chi$ is fixed by the equations of state, we take it to be a constant in the following analysis for simplicity. This relation implies the late-time scaling
$\gamma_s \sim (\beta/\alpha)(\chi /8)\tau^{-1}$. 
Inserting this scaling into Eq.~(\ref{eq:con_phi}) shows that, for $\beta/\alpha>1$, the orbital angular-momentum exchange term (encoded in the source $\phi^{\mu\nu}$) decreases more slowly than the typical late-time scaling $\tau^{-1}$. Consequently, the source term remains sufficiently large at late times and can counteract the expansion-driven dilution of the spin density. 
As a result, the spin density can become nearly constant or even increase with $\tau$, consistent with the nontrivial static or growing behavior found in Eq.~(\ref{eq:S_beta_alpha_01}).
In contrast, if $\beta/\alpha<1$, then $\gamma_s$ (and thus $\phi^{\mu\nu}$) is suppressed, so the source term becomes subleading and the spin density follows the usual power-law decay at late times.

\section{Conclusion}
\label{sec:summary}

In this work, we have investigated the late-time asymptotic solutions and attractor structure of minimal causal spin hydrodynamics in Gubser flow. We first derive the governing differential equation for the spin density $\hat{S}$ in Eq.~\eqref{eq:Eq_S} and introduce the auxiliary function $f(w)$ in Eq.~\eqref{eq:f_S}. We then obtain the large $w$ asymptotic solutions of $f(w)$, summarized in Table~\ref{tab:Solutions_f}. Next, we present numerical solutions for $f(w)$ for different parameter choices, shown in Fig.~\ref{fig:attractor}. We identify both attractors and repellers in Fig.~\ref{fig:attractor} and analyze their properties.
Finally, we map the asymptotic solutions back to the spin density in flat Minkowski space with coordinates $(t,x,y,z)$, and we delineate the parameter regions associated with different decay behaviors in Fig.~\ref{fig:region}. In particular, by considering the two late-time regimes, case (i) $\Delta_{1}>\Delta_{2}$ and case (ii) $\Delta_{1}=\Delta_{2}$, we identify parameter regions in which the spin density exhibits a non-vanishing power-law decay rather than an exponential decay. This result indicates that, for suitable parameter choices, the spin density can evolve as a hydrodynamic mode governed by the asymptotic scaling of the underlying flow, rather than as a rapidly damped non-hydrodynamic variable.

\begin{acknowledgments}
This work is supported in part by the National Key Research and Development
Program of China under Contract No.~2022YFA1605500, by the Chinese
Academy of Sciences (CAS) under Grant No. YSBR-088 and by National Natural Science Foundation of China (NSFC) under Grants No.~12135011. D.-L. Wang is supported in part by the National Natural Science Foundation of China under Grant No.~125B2110.
\end{acknowledgments}

\bibliographystyle{h-physrev}
\bibliography{refs}

\begin{thebibliography}{100}

\bibitem{Liang:2004ph}
Z.-T. Liang and X.-N. Wang,
\newblock Phys. Rev. Lett. {\bf 94}, 102301 (2005), nucl-th/0410079,
\newblock [Erratum: Phys.Rev.Lett. 96, 039901 (2006)].

\bibitem{Liang:2004xn}
Z.-T. Liang and X.-N. Wang,
\newblock Phys. Lett. B {\bf 629}, 20 (2005), nucl-th/0411101.

\bibitem{STAR:2017ckg}
STAR, L.~Adamczyk {\em et~al.},
\newblock Nature {\bf 548}, 62 (2017), 1701.06657.

\bibitem{STAR:2018gyt}
STAR, J.~Adam {\em et~al.},
\newblock Phys. Rev. C {\bf 98}, 014910 (2018), 1805.04400.

\bibitem{STAR:2020xbm}
STAR, J.~Adam {\em et~al.},
\newblock Phys. Rev. Lett. {\bf 126}, 162301 (2021), 2012.13601.

\bibitem{STAR:2021beb}
STAR, M.~S. Abdallah {\em et~al.},
\newblock Phys. Rev. C {\bf 104}, L061901 (2021), 2108.00044.

\bibitem{STAR:2023nvo}
STAR, M.~I. Abdulhamid {\em et~al.},
\newblock Phys. Rev. C {\bf 108}, 014910 (2023), 2305.08705.

\bibitem{Karpenko:2016jyx}
I.~Karpenko and F.~Becattini,
\newblock Eur. Phys. J. C {\bf 77}, 213 (2017), 1610.04717.

\bibitem{Li:2017slc}
H.~Li, L.-G. Pang, Q.~Wang, and X.-L. Xia,
\newblock Phys. Rev. C {\bf 96}, 054908 (2017), 1704.01507.

\bibitem{Sun:2017xhx}
Y.~Sun and C.~M. Ko,
\newblock Phys. Rev. C {\bf 96}, 024906 (2017), 1706.09467.

\bibitem{Wei:2018zfb}
D.-X. Wei, W.-T. Deng, and X.-G. Huang,
\newblock Phys. Rev. C {\bf 99}, 014905 (2019), 1810.00151.

\bibitem{Vitiuk:2019rfv}
O.~Vitiuk, L.~V. Bravina, and E.~E. Zabrodin,
\newblock Phys. Lett. B {\bf 803}, 135298 (2020), 1910.06292.

\bibitem{Fu:2020oxj}
B.~Fu, K.~Xu, X.-G. Huang, and H.~Song,
\newblock Phys. Rev. C {\bf 103}, 024903 (2021), 2011.03740.

\bibitem{Lei:2021mvp}
A.~Lei, D.~Wang, D.-M. Zhou, B.-H. Sa, and L.~P. Csernai,
\newblock Phys. Rev. C {\bf 104}, 054903 (2021), 2110.13485.

\bibitem{Ryu:2021lnx}
S.~Ryu, V.~Jupic, and C.~Shen,
\newblock Phys. Rev. C {\bf 104}, 054908 (2021), 2106.08125.

\bibitem{HADES:2022enx}
HADES, R.~Abou~Yassine {\em et~al.},
\newblock Phys. Lett. B {\bf 835}, 137506 (2022), 2207.05160.

\bibitem{Guo:2021udq}
Y.~Guo, J.~Liao, E.~Wang, H.~Xing, and H.~Zhang,
\newblock Phys. Rev. C {\bf 104}, L041902 (2021), 2105.13481.

\bibitem{Deng:2020ygd}
X.-G. Deng, X.-G. Huang, Y.-G. Ma, and S.~Zhang,
\newblock Phys. Rev. C {\bf 101}, 064908 (2020), 2001.01371.

\bibitem{Deng:2021miw}
X.-G. Deng, X.-G. Huang, and Y.-G. Ma,
\newblock (2021), 2109.09956.

\bibitem{Sun:2025oib}
K.-J. Sun {\em et~al.},
\newblock Phys. Rev. Lett. {\bf 134}, 022301 (2025).

\bibitem{Liu:2025kpp}
D.-N. Liu {\em et~al.},
\newblock (2025), 2508.12193.

\bibitem{Zheng:2025ngn}
Y.-P. Zheng {\em et~al.},
\newblock (2025), 2509.15286.

\bibitem{Xu:2026hxz}
J.~Xu,
\newblock (2026), 2602.23793.

\bibitem{Liu:2021uhn}
S.~Y.~F. Liu and Y.~Yin,
\newblock JHEP {\bf 07}, 188 (2021), 2103.09200.

\bibitem{Fu:2021pok}
B.~Fu, S.~Y.~F. Liu, L.~Pang, H.~Song, and Y.~Yin,
\newblock Phys. Rev. Lett. {\bf 127}, 142301 (2021), 2103.10403.

\bibitem{Becattini:2021suc}
F.~Becattini, M.~Buzzegoli, and A.~Palermo,
\newblock Phys. Lett. B {\bf 820}, 136519 (2021), 2103.10917.

\bibitem{Becattini:2021iol}
F.~Becattini, M.~Buzzegoli, G.~Inghirami, I.~Karpenko, and A.~Palermo,
\newblock Phys. Rev. Lett. {\bf 127}, 272302 (2021), 2103.14621.

\bibitem{Hidaka:2017auj}
Y.~Hidaka, S.~Pu, and D.-L. Yang,
\newblock Phys. Rev. D {\bf 97}, 016004 (2018), 1710.00278.

\bibitem{Yi:2021ryh}
C.~Yi, S.~Pu, and D.-L. Yang,
\newblock Phys. Rev. C {\bf 104}, 064901 (2021), 2106.00238.

\bibitem{Yi:2021unq}
C.~Yi, S.~Pu, J.-H. Gao, and D.-L. Yang,
\newblock Phys. Rev. C {\bf 105}, 044911 (2022), 2112.15531.

\bibitem{Florkowski:2021xvy}
W.~Florkowski, A.~Kumar, A.~Mazeliauskas, and R.~Ryblewski,
\newblock Phys. Rev. C {\bf 105}, 064901 (2022), 2112.02799.

\bibitem{Alzhrani:2022dpi}
S.~Alzhrani, S.~Ryu, and C.~Shen,
\newblock Phys. Rev. C {\bf 106}, 014905 (2022), 2203.15718.

\bibitem{Palermo:2022lvh}
A.~Palermo, F.~Becattini, M.~Buzzegoli, G.~Inghirami, and I.~Karpenko,
\newblock EPJ Web Conf. {\bf 276}, 01026 (2023), 2208.09874.

\bibitem{Buzzegoli:2022fxu}
M.~Buzzegoli, F.~Becattini, G.~Inghirami, I.~Karpenko, and A.~Palermo,
\newblock Acta Phys. Polon. Supp. {\bf 16}, 39 (2023), 2208.04449.

\bibitem{Wu:2022mkr}
X.-Y. Wu, C.~Yi, G.-Y. Qin, and S.~Pu,
\newblock Phys. Rev. C {\bf 105}, 064909 (2022), 2204.02218.

\bibitem{Yi:2023tgg}
C.~Yi {\em et~al.},
\newblock Phys. Rev. C {\bf 109}, L011901 (2024), 2304.08777.

\bibitem{Wu:2023tku}
X.-Y. Wu, C.~Yi, G.-Y. Qin, and S.~Pu,
\newblock {Global and local polarization of $\Lambda$ hyperons across RHIC-BES
  energies},
\newblock in {\em {30th International Conference on Ultrarelativstic
  Nucleus-Nucleus Collisions}}, 2023, 2312.09068.

\bibitem{Palermo:2024tza}
A.~Palermo, E.~Grossi, I.~Karpenko, and F.~Becattini,
\newblock Eur. Phys. J. C {\bf 84}, 920 (2024), 2404.14295.

\bibitem{Fu:2022myl}
B.~Fu, L.~Pang, H.~Song, and Y.~Yin,
\newblock (2022), 2201.12970.

\bibitem{Ivanov:2022geb}
Y.~B. Ivanov and A.~A. Soldatov,
\newblock Pisma Zh. Eksp. Teor. Fiz. {\bf 116}, 137 (2022), 2206.06927.

\bibitem{Fang:2024vds}
S.~Fang and S.~Pu,
\newblock Phys. Rev. D {\bf 111}, 034015 (2025), 2408.09877.

\bibitem{Wang:2025mfz}
J.-R. Wang, S.~Fang, D.-L. Yang, and S.~Pu,
\newblock (2025), 2507.15238.

\bibitem{Fang:2022ttm}
S.~Fang, S.~Pu, and D.-L. Yang,
\newblock Phys. Rev. D {\bf 106}, 016002 (2022), 2204.11519.

\bibitem{Fang:2025pzy}
S.~Fang, S.~Pu, and D.-L. Yang,
\newblock Phys. Rev. D {\bf 112}, 014038 (2025), 2503.13320.

\bibitem{Yi:2024kwu}
C.~Yi, X.-Y. Wu, J.~Zhu, S.~Pu, and G.-Y. Qin,
\newblock Phys. Rev. C {\bf 111}, 044901 (2025), 2408.04296.

\bibitem{CMS:2025nqr}
CMS, A.~Hayrapetyan {\em et~al.},
\newblock Phys. Rev. Lett. {\bf 135}, 132301 (2025), 2502.07898.

\bibitem{Yi:2025anh}
C.~Yi, S.~Fang, D.-L. Wang, and S.~Pu,
\newblock {Local spin polarization of $\Lambda$ hyperons and its interaction
  corrections},
\newblock in {\em {31st International Conference on Ultra-relativistic
  Nucleus-Nucleus Collisions}}, 2025, 2509.00377.

\bibitem{Yi:2025sgk}
C.~Yi, X.-Y. Wu, J.~Zhu, S.~Pu, and G.-Y. Qin,
\newblock {Hydrodynamic effects on spin polarization along the beam direction
  in Au+Au and p+Pb collisions},
\newblock in {\em {31st International Conference on Ultra-relativistic
  Nucleus-Nucleus Collisions}}, 2025, 2509.00380.

\bibitem{Gao:2020vbh}
J.-H. Gao, G.-L. Ma, S.~Pu, and Q.~Wang,
\newblock Nucl. Sci. Tech. {\bf 31}, 90 (2020), 2005.10432.

\bibitem{Becattini:2022zvf}
F.~Becattini,
\newblock Rept. Prog. Phys. {\bf 85}, 122301 (2022), 2204.01144.

\bibitem{Becattini:2024uha}
F.~Becattini {\em et~al.},
\newblock (2024), 2402.04540.

\bibitem{Niida:2024ntm}
T.~Niida and S.~A. Voloshin,
\newblock Int. J. Mod. Phys. E {\bf 33}, 2430010 (2024), 2404.11042.

\bibitem{Gao:2012ix}
J.-H. Gao, Z.-T. Liang, S.~Pu, Q.~Wang, and X.-N. Wang,
\newblock Phys. Rev. Lett. {\bf 109}, 232301 (2012), 1203.0725.

\bibitem{Chen:2012ca}
J.-W. Chen, S.~Pu, Q.~Wang, and X.-N. Wang,
\newblock Phys. Rev. Lett. {\bf 110}, 262301 (2013), 1210.8312.

\bibitem{Hidaka:2016yjf}
Y.~Hidaka, S.~Pu, and D.-L. Yang,
\newblock Phys. Rev. D {\bf 95}, 091901 (2017), 1612.04630.

\bibitem{Gao:2019znl}
J.-H. Gao and Z.-T. Liang,
\newblock Phys. Rev. D {\bf 100}, 056021 (2019), 1902.06510.

\bibitem{Weickgenannt:2019dks}
N.~Weickgenannt, X.-L. Sheng, E.~Speranza, Q.~Wang, and D.~H. Rischke,
\newblock Phys. Rev. D {\bf 100}, 056018 (2019), 1902.06513.

\bibitem{Liu:2020flb}
Y.-C. Liu, K.~Mameda, and X.-G. Huang,
\newblock Chin. Phys. C {\bf 44}, 094101 (2020), 2002.03753,
\newblock [Erratum: Chin.Phys.C 45, 089001 (2021)].

\bibitem{Weickgenannt:2020aaf}
N.~Weickgenannt, E.~Speranza, X.-l. Sheng, Q.~Wang, and D.~H. Rischke,
\newblock Phys. Rev. Lett. {\bf 127}, 052301 (2021), 2005.01506.

\bibitem{Weickgenannt:2021cuo}
N.~Weickgenannt, E.~Speranza, X.-l. Sheng, Q.~Wang, and D.~H. Rischke,
\newblock Phys. Rev. D {\bf 104}, 016022 (2021), 2103.04896.

\bibitem{Sheng:2021kfc}
X.-L. Sheng, N.~Weickgenannt, E.~Speranza, D.~H. Rischke, and Q.~Wang,
\newblock Phys. Rev. D {\bf 104}, 016029 (2021), 2103.10636.

\bibitem{Hidaka:2022dmn}
Y.~Hidaka, S.~Pu, Q.~Wang, and D.-L. Yang,
\newblock Prog. Part. Nucl. Phys. {\bf 127}, 103989 (2022), 2201.07644.

\bibitem{Dong:2022yzt}
W.-B. Dong, Y.-L. Yin, and Q.~Wang,
\newblock Phys. Rev. C {\bf 106}, 054909 (2022), 2209.12402.

\bibitem{Dong:2023cng}
W.-B. Dong, Y.-L. Yin, X.-L. Sheng, S.-Z. Yang, and Q.~Wang,
\newblock Phys. Rev. D {\bf 109}, 056025 (2024), 2311.18400.

\bibitem{Fang:2023bbw}
S.~Fang, S.~Pu, and D.-L. Yang,
\newblock Phys. Rev. D {\bf 109}, 034034 (2024), 2311.15197.

\bibitem{Yin:2024dnu}
Y.-L. Yin, W.-B. Dong, J.-Y. Pang, S.~Pu, and Q.~Wang,
\newblock (2024), 2402.03672.

\bibitem{Bhadury:2025boe}
S.~Bhadury, Z.~Drogosz, W.~Florkowski, S.~K. Kar, and V.~Mykhaylova,
\newblock (2025), 2505.02657.

\bibitem{Montenegro:2017rbu}
D.~Montenegro, L.~Tinti, and G.~Torrieri,
\newblock Phys. Rev. D {\bf 96}, 056012 (2017), 1701.08263,
\newblock [Addendum: Phys.Rev.D 96, 079901 (2017)].

\bibitem{Florkowski:2017ruc}
W.~Florkowski, B.~Friman, A.~Jaiswal, and E.~Speranza,
\newblock Phys. Rev. C {\bf 97}, 041901 (2018), 1705.00587.

\bibitem{Florkowski:2018fap}
W.~Florkowski, A.~Kumar, and R.~Ryblewski,
\newblock Prog. Part. Nucl. Phys. {\bf 108}, 103709 (2019), 1811.04409.

\bibitem{Hattori:2019lfp}
K.~Hattori, M.~Hongo, X.-G. Huang, M.~Matsuo, and H.~Taya,
\newblock Phys. Lett. B {\bf 795}, 100 (2019), 1901.06615.

\bibitem{Fukushima:2020ucl}
K.~Fukushima and S.~Pu,
\newblock Phys. Lett. B {\bf 817}, 136346 (2021), 2010.01608.

\bibitem{Becattini:2018duy}
F.~Becattini, W.~Florkowski, and E.~Speranza,
\newblock Phys. Lett. B {\bf 789}, 419 (2019), 1807.10994.

\bibitem{Hu:2022azy}
J.~Hu,
\newblock Phys. Rev. C {\bf 107}, 024915 (2023), 2209.10979.

\bibitem{Dey:2024cwo}
S.~Dey and A.~Das,
\newblock Phys. Rev. D {\bf 111}, 074037 (2025), 2410.04141.

\bibitem{Tiwari:2024trl}
A.~Tiwari and B.~K. Patra,
\newblock Phys. Rev. D {\bf 112}, 036014 (2025), 2408.11514.

\bibitem{She:2024rnx}
D.~She, Y.-W. Qiu, and D.~Hou,
\newblock Phys. Rev. D {\bf 111}, 036027 (2025), 2410.15142.

\bibitem{Florkowski:2018myy}
W.~Florkowski, E.~Speranza, and F.~Becattini,
\newblock Acta Phys. Polon. B {\bf 49}, 1409 (2018), 1803.11098.

\bibitem{Bhadury:2020puc}
S.~Bhadury, W.~Florkowski, A.~Jaiswal, A.~Kumar, and R.~Ryblewski,
\newblock Phys. Lett. B {\bf 814}, 136096 (2021), 2002.03937.

\bibitem{Shi:2020htn}
S.~Shi, C.~Gale, and S.~Jeon,
\newblock Phys. Rev. C {\bf 103}, 044906 (2021), 2008.08618.

\bibitem{Speranza:2020ilk}
E.~Speranza and N.~Weickgenannt,
\newblock Eur. Phys. J. A {\bf 57}, 155 (2021), 2007.00138.

\bibitem{Bhadury:2020cop}
S.~Bhadury, W.~Florkowski, A.~Jaiswal, A.~Kumar, and R.~Ryblewski,
\newblock Phys. Rev. D {\bf 103}, 014030 (2021), 2008.10976.

\bibitem{Singh:2020rht}
R.~Singh, G.~Sophys, and R.~Ryblewski,
\newblock Phys. Rev. D {\bf 103}, 074024 (2021), 2011.14907.

\bibitem{Peng:2021ago}
H.-H. Peng, J.-J. Zhang, X.-L. Sheng, and Q.~Wang,
\newblock Chin. Phys. Lett. {\bf 38}, 116701 (2021), 2107.00448.

\bibitem{Weickgenannt:2022zxs}
N.~Weickgenannt, D.~Wagner, E.~Speranza, and D.~H. Rischke,
\newblock Phys. Rev. D {\bf 106}, 096014 (2022), 2203.04766.

\bibitem{Weickgenannt:2022jes}
N.~Weickgenannt, D.~Wagner, and E.~Speranza,
\newblock Phys. Rev. D {\bf 105}, 116026 (2022), 2204.01797.

\bibitem{Weickgenannt:2022qvh}
N.~Weickgenannt, D.~Wagner, E.~Speranza, and D.~H. Rischke,
\newblock Phys. Rev. D {\bf 106}, L091901 (2022), 2208.01955.

\bibitem{Bhadury:2022ulr}
S.~Bhadury, W.~Florkowski, A.~Jaiswal, A.~Kumar, and R.~Ryblewski,
\newblock Phys. Rev. Lett. {\bf 129}, 192301 (2022), 2204.01357.

\bibitem{Wagner:2023cct}
D.~Wagner, N.~Weickgenannt, and E.~Speranza,
\newblock Phys. Rev. D {\bf 108}, 116017 (2023), 2306.05936.

\bibitem{Florkowski:2024bfw}
W.~Florkowski and M.~Hontarenko,
\newblock Phys. Rev. Lett. {\bf 134}, 082302 (2025), 2405.03263.

\bibitem{Wagner:2024fhf}
D.~Wagner, M.~Shokri, and D.~H. Rischke,
\newblock (2024), 2405.00533.

\bibitem{Wagner:2024fry}
D.~Wagner,
\newblock Phys. Rev. D {\bf 111}, 016008 (2025), 2409.07143.

\bibitem{Daher:2025pfq}
A.~Daher, X.-L. Sheng, D.~Wagner, and F.~Becattini,
\newblock Phys. Rev. D {\bf 112}, 094020 (2025), 2503.03713.

\bibitem{Gallegos:2020otk}
A.~D. Gallegos and U.~G\"ursoy,
\newblock JHEP {\bf 11}, 151 (2020), 2004.05148.

\bibitem{Li:2020eon}
S.~Li, M.~A. Stephanov, and H.-U. Yee,
\newblock Phys. Rev. Lett. {\bf 127}, 082302 (2021), 2011.12318.

\bibitem{Gallegos:2021bzp}
A.~D. Gallegos, U.~G\"ursoy, and A.~Yarom,
\newblock SciPost Phys. {\bf 11}, 041 (2021), 2101.04759.

\bibitem{She:2021lhe}
D.~She, A.~Huang, D.~Hou, and J.~Liao,
\newblock Sci. Bull. {\bf 67}, 2265 (2022), 2105.04060.

\bibitem{Hongo:2021ona}
M.~Hongo, X.-G. Huang, M.~Kaminski, M.~Stephanov, and H.-U. Yee,
\newblock JHEP {\bf 11}, 150 (2021), 2107.14231.

\bibitem{Biswas:2023qsw}
R.~Biswas, A.~Daher, A.~Das, W.~Florkowski, and R.~Ryblewski,
\newblock (2023), 2304.01009.

\bibitem{Fang:2025aig}
S.~Fang, K.~Fukushima, S.~Pu, and D.-L. Wang,
\newblock (2025), 2506.20698.

\bibitem{Zhang:2026zee}
Z.-H. Zhang, X.-H. Lv, and X.-G. Huang,
\newblock (2026), 2603.17794.

\bibitem{Montenegro:2017lvf}
D.~Montenegro, L.~Tinti, and G.~Torrieri,
\newblock Phys. Rev. D {\bf 96}, 076016 (2017), 1703.03079.

\bibitem{Shi:2023sxh}
P.~Shi and H.~Xu-Guang,
\newblock Acta Phys. Sin. {\bf 72}, 071202 (2023).

\bibitem{Huang:2024ffg}
X.-G. Huang,
\newblock Nucl. Sci. Tech. {\bf 36}, 208 (2025), 2411.11753.

\bibitem{Xie:2023gbo}
X.-Q. Xie, D.-L. Wang, C.~Yang, and S.~Pu,
\newblock Phys. Rev. D {\bf 108}, 094031 (2023), 2306.13880.

\bibitem{Sarwar:2022yzs}
G.~Sarwar, M.~Hasanujjaman, J.~R. Bhatt, H.~Mishra, and J.-e. Alam,
\newblock Phys. Rev. D {\bf 107}, 054031 (2023), 2209.08652.

\bibitem{Daher:2022wzf}
A.~Daher, A.~Das, and R.~Ryblewski,
\newblock Phys. Rev. D {\bf 107}, 054043 (2023), 2209.10460.

\bibitem{Daher:2024bah}
A.~Daher, W.~Florkowski, R.~Ryblewski, and F.~Taghinavaz,
\newblock (2024), 2403.04711.

\bibitem{Denicol:2008ha}
G.~S. Denicol, T.~Kodama, T.~Koide, and P.~Mota,
\newblock J. Phys. G {\bf 35}, 115102 (2008), 0807.3120.

\bibitem{Pu:2009fj}
S.~Pu, T.~Koide, and D.~H. Rischke,
\newblock Phys. Rev. D {\bf 81}, 114039 (2010), 0907.3906.

\bibitem{Kovtun:2019hdm}
P.~Kovtun,
\newblock JHEP {\bf 10}, 034 (2019), 1907.08191.

\bibitem{Heller:2022ejw}
M.~P. Heller, A.~Serantes, M.~Spali\'nski, and B.~Withers,
\newblock Phys. Rev. Lett. {\bf 130}, 261601 (2023), 2212.07434.

\bibitem{Gavassino:2023myj}
L.~Gavassino,
\newblock Phys. Lett. B {\bf 840}, 137854 (2023), 2301.06651.

\bibitem{Gavassino:2023mad}
L.~Gavassino, M.~M. Disconzi, and J.~Noronha,
\newblock Phys. Rev. Lett. {\bf 132}, 162301 (2024), 2307.05987.

\bibitem{Wang:2023csj}
D.-L. Wang and S.~Pu,
\newblock Phys. Rev. D {\bf 109}, L031504 (2024), 2309.11708.

\bibitem{Hoult:2023clg}
R.~E. Hoult and P.~Kovtun,
\newblock Phys. Rev. D {\bf 109}, 046018 (2024), 2309.11703.

\bibitem{Ren:2024pur}
X.~Ren, C.~Yang, D.-L. Wang, and S.~Pu,
\newblock (2024), 2405.03105.

\bibitem{Gavassino:2021kjm}
L.~Gavassino, M.~Antonelli, and B.~Haskell,
\newblock Phys. Rev. Lett. {\bf 128}, 010606 (2022), 2105.14621.

\bibitem{Dey:2023hft}
S.~Dey, W.~Florkowski, A.~Jaiswal, and R.~Ryblewski,
\newblock (2023), 2303.05271.

\bibitem{Buzzegoli:2021wlg}
M.~Buzzegoli,
\newblock Phys. Rev. C {\bf 105}, 044907 (2022), 2109.12084.

\bibitem{Buzzegoli:2024mra}
M.~Buzzegoli and A.~Palermo,
\newblock Phys. Rev. Lett. {\bf 133}, 262301 (2024), 2407.14345.

\bibitem{Becattini:2025twu}
F.~Becattini and C.~Hoyos,
\newblock (2025), 2507.09249.

\bibitem{Becattini:2023ouz}
F.~Becattini, A.~Daher, and X.-L. Sheng,
\newblock Phys. Lett. B {\bf 850}, 138533 (2024), 2309.05789.

\bibitem{Becattini:2025oyi}
F.~Becattini and R.~Singh,
\newblock Eur. Phys. J. C {\bf 85}, 1338 (2025), 2506.20681.

\bibitem{Ambrus:2025dca}
V.~E. Ambrus and A.~Geci{\'c},
\newblock (2025), 2509.17640.

\bibitem{Armas:2026bmw}
J.~Armas and A.~Jain,
\newblock (2026), 2601.14421.

\bibitem{Wang:2021ngp}
D.-L. Wang, S.~Fang, and S.~Pu,
\newblock Phys. Rev. D {\bf 104}, 114043 (2021), 2107.11726.

\bibitem{Wang:2021wqq}
D.-L. Wang, X.-Q. Xie, S.~Fang, and S.~Pu,
\newblock Phys. Rev. D {\bf 105}, 114050 (2022), 2112.15535.

\bibitem{Wang:2024afv}
D.-L. Wang, L.~Yan, and S.~Pu,
\newblock Phys. Rev. D {\bf 111}, 034033 (2025), 2408.03781.

\bibitem{Heller:2015dha}
M.~P. Heller and M.~Spalinski,
\newblock Phys. Rev. Lett. {\bf 115}, 072501 (2015), 1503.07514.

\bibitem{Romatschke:2017vte}
P.~Romatschke,
\newblock Phys. Rev. Lett. {\bf 120}, 012301 (2018), 1704.08699.

\bibitem{Blaizot:2017ucy}
J.-P. Blaizot and L.~Yan,
\newblock Phys. Lett. B {\bf 780}, 283 (2018), 1712.03856.

\bibitem{Denicol:2017lxn}
G.~S. Denicol and J.~Noronha,
\newblock Phys. Rev. D {\bf 97}, 056021 (2018), 1711.01657.

\bibitem{Spalinski:2017mel}
M.~Spali\'nski,
\newblock Phys. Lett. B {\bf 776}, 468 (2018), 1708.01921.

\bibitem{Strickland:2018ayk}
M.~Strickland,
\newblock JHEP {\bf 12}, 128 (2018), 1809.01200.

\bibitem{Blaizot:2019scw}
J.-P. Blaizot and L.~Yan,
\newblock Annals Phys. {\bf 412}, 167993 (2020), 1904.08677.

\bibitem{Brewer:2019oha}
J.~Brewer, L.~Yan, and Y.~Yin,
\newblock Phys. Lett. B {\bf 816}, 136189 (2021), 1910.00021.

\bibitem{Almaalol:2020rnu}
D.~Almaalol, A.~Kurkela, and M.~Strickland,
\newblock Phys. Rev. Lett. {\bf 125}, 122302 (2020), 2004.05195.

\bibitem{Blaizot:2020gql}
J.-P. Blaizot and L.~Yan,
\newblock Phys. Lett. B {\bf 820}, 136478 (2021), 2006.08815.

\bibitem{Heller:2020anv}
M.~P. Heller, R.~Jefferson, M.~Spali\'nski, and V.~Svensson,
\newblock Phys. Rev. Lett. {\bf 125}, 132301 (2020), 2003.07368.

\bibitem{Heller:2020uuy}
M.~P. Heller, A.~Serantes, M.~Spali\'nski, V.~Svensson, and B.~Withers,
\newblock Phys. Rev. D {\bf 104}, 066002 (2021), 2007.05524.

\bibitem{Heller:2021oxl}
M.~P. Heller, A.~Serantes, M.~Spali\'nski, V.~Svensson, and B.~Withers,
\newblock Phys. Rev. Lett. {\bf 128}, 122302 (2022), 2110.07621.

\bibitem{Blaizot:2021cdv}
J.-P. Blaizot and L.~Yan,
\newblock Phys. Rev. C {\bf 104}, 055201 (2021), 2106.10508.

\bibitem{Chen:2022ryi}
Z.~Chen, D.~Teaney, and L.~Yan,
\newblock Phys. Rev. C {\bf 108}, 064911 (2023), 2206.12778.

\bibitem{Gubser:2010ze}
S.~S. Gubser,
\newblock Phys. Rev. D {\bf 82}, 085027 (2010), 1006.0006.

\bibitem{Gubser:2010ui}
S.~S. Gubser and A.~Yarom,
\newblock Nucl. Phys. B {\bf 846}, 469 (2011), 1012.1314.

\bibitem{Strickland:2017kux}
M.~Strickland, J.~Noronha, and G.~Denicol,
\newblock Phys. Rev. D {\bf 97}, 036020 (2018), 1709.06644.

\bibitem{Denicol:2018pak}
G.~S. Denicol and J.~Noronha,
\newblock Phys. Rev. D {\bf 99}, 116004 (2019), 1804.04771.

\end{thebibliography}
%%%%%%%%%%%%%%%%%%%

\end{document}